%% file: vrlecture_revised.tex
\documentclass[a4paper,10pt]{article}

\usepackage{graphicx}
\usepackage{latexsym}
\usepackage{amsfonts,amsmath,amssymb}
\usepackage{color}

\def\g{\sqrt{-g}\,}

\def\rn{{Reissner-Nordstrom}}

\newcommand{\LL}{Lanczos-Lovelock}

\newcommand{\Cal}[1]{\ensuremath{\mathcal{#1}}}
\newcommand{\ph}[1]{\phantom{#1}}
\newcommand{\D}{\ensuremath{\nabla}}
 
\newcommand{\Alt}[6]{\ensuremath{\delta^{#1 #2 ... #3}_{#4 #5
      ... #6}}} 
       
\newcommand{\Riem}[4]{\ensuremath{R^{#1 #2}_{#3 #4}}}
\newcommand{\LDm}{\ensuremath{L_{(m)}}}
\newcommand{\sD}[1]{\sum_{m=1}^{K}{#1}}
\newcommand{\dV}{\ensuremath{\partial\Cal{V}}}

\newcommand{\cc}{cosmological constant}

\def\eq#1{{Eq.~(\ref{#1})}}

\newcommand{\ha}{\vskip0.2cm\noindent\textbf{Harold:\ }}
\newcommand{\me}{\vskip0.2cm\noindent\textbf{Me:\ }}

\title{A Dialogue on the Nature of Gravity}
\author{T. Padmanabhan\\
IUCAA, Pune University Campus,\\
Ganeshkhind, Pune 411007, INDIA.}

\date{ }

\begin{document}

\maketitle

\begin{abstract}
I describe the conceptual and mathematical basis of an approach which describes gravity as an emergent phenomenon. Combining the principle of equivalence and the principle of general
covariance with known properties of local Rindler horizons, perceived by 
observers accelerated with respect to local inertial frames, one can provide a thermodynamic re-interpretation of the field equations describing gravity in any diffeomorphism
invariant theory. 
This fact, in turn, leads us to the possibility of deriving the field equations of gravity by maximising a suitably defined entropy functional, without using the metric tensor as a dynamical variable.
The approach synthesizes concepts from quantum theory, thermodynamics and gravity leading to a fresh perspective on the nature of gravity.
The description is presented here in the form of a dialogue, thereby addressing several frequently-asked-questions.

\end{abstract}

\section{What is it all about?}

\ha \footnote{Harold was  a very useful creation originally due to Julian Schwinger \cite{julian}
and stands for Hypothetically Alert Reader Of Limitless Dedication. In the present context, I think of Harold as Hypothetically Alert Relativist  Open to Logical Discussions.} 
For quite sometime now, you have been talking about `gravity being an emergent phenomenon' and a `thermodynamic perspective on gravity'. This is  quite different from the conventional point of view in which gravity is a fundamental interaction and spacetime thermodynamics of, say, black holes is a particular result which can be derived in a specific context. Honestly, while I find your papers fascinating I am not clear about the broad picture you are trying to convey. Maybe you could begin by clarifying what this is all about, before we plunge into the details ? What is the roadmap, so to speak ?

\me To begin with, I will show you that the equations motion describing gravity in  \textit{any diffeomorphism invariant theory} can be given \cite{tp09papers} a  suggestive thermodynamic re-interpretation (Sections \ref{sec:lro}, \ref{sec:reinterpret}). Second, taking a cue from this, I can formulate a variational principle for a suitably defined entropy functional --- involving both gravity and matter --- which will lead to the field equations of gravity \cite{aseementropy,tpbojo} without varying the metric tensor as a dynamical variable (Section \ref{sec:eqnnewvar}).

\ha Suppose I have an action for gravity plus matter (in $D$ dimensions)
\begin{equation}
A=\int d^Dx \sqrt{-g}\left[L(R^{ab}_{cd}, g^{ab})+L_{matt}(g^{ab},q_A)\right]
\end{equation}  
where $ L$ is any scalar built from metric and curvature  and $L_{matt}$ is the matter Lagrangian depending on the metric and some matter variables $q_A$. (I will assume $L$ does not involve derivatives of curvature tensor, to simplify the discussion.) If I vary $g^{ab}$ in the action I will get some  equations of motion (see e.g. Refs. \cite{ayan,mohut}), say,  $2E_{ab}=T_{ab}$ where $E_{ab}$ is \footnote{The signature is - + + + and Latin letters cover spacetime indices while Greek letters run over space indices.} 
\begin{equation}
E_{ab}=P_a^{\phantom{a} cde} R_{bcde} - 2 \nabla^c \nabla^d P_{acdb} - \frac{1}{2} L g_{ab};\quad
P^{abcd} \equiv \frac{\partial L}{\partial R_{abcd}}
\label{genEab}
\end{equation} 
Now, you are telling me that (i) you can give a thermodynamic interpretation 
to the equation 
$2E_{ab}=T_{ab}$ just because it comes from a scalar Lagrangian and (ii) you can also derive it from an entropy maximisation principle. I admit it is  fascinating. But why should I take this  approach as more fundamental, conceptually, than the good old way of just varying the total Lagrangian $L+L_{matt}$
and getting $2E_{ab}=T_{ab}$ ? Why is it more than a curiosity ?

\me That brings me to the third aspect of the formulation which I will discuss towards the end (Section \ref{sec:comaparison}). In my approach, I can provide a natural explanation to several puzzling aspects of gravity and horizon thermodynamics all of which have to be thought of as mere algebraic accidents in the conventional approach you mentioned. Let me give an analogy. In Newtonian gravity, the fact that inertial mass is equal to the gravitational mass is an algebraic accident without any fundamental explanation. But in a geometrical theory of gravity based on principle of equivalence, this fact finds a natural explanation. Similarly, I think we can make progress by identifying key facts  which have no explanation in the conventional approach and providing them  a natural explanation from a different perspective. You will also see that this approach connects up several pieces of conventional theory  in an elegant manner.

\ha Your ideas also seem to be quite different from other works which describe gravity as an emergent phenomena \cite{analog}.
Can you explain \textit{your} motivation?

\me 
Yes. The original inspiration for my work, as for many others, comes from the old idea of Sakharov \cite{sakharov}
which attempted  to describe spacetime dynamics as akin to the theory of elasticity. There are two crucial differences between my approach and many other ones. 

To begin with,
I realized that the thermodynamic description transcends Einstein's general relativity and can incorporate a much wider class of theories --- this was first pointed out in ref. \cite{paris} and elaborated in several of my papers --- while many other approaches concentrated on just Einstein's theory. In fact, many other approaches use techniques strongly linked to Einstein's theory -- like for example, Raychaudhuri equation to study rate of change of horizon area, which is difficult to  generalize to theories in which the horizon entropy is \textit{not} proportional to horizon area. I use more general techniques. 

Second, I work at the level of action principle and its symmetries to a large extent  so I have a handle on the off-shell structure of the theory; in fact, much of the thermodynamic interpretation in my approach is closely linked to the structure of action functional (like e.g., the existence of surface term in action, holographic nature etc.) for gravitational theories. This link is central to me while it is not taken into account in any other approach.

\ha So essentially you are claiming that the thermodynamics of horizons is more central than the dynamics of the gravitational field while the conventional view is probably the other way around. Why do you stress the thermal aspects of horizons so much ? Can you give a motivation ?
 
\me Because thermal phenomenon is a window to microstructure! Let me explain. We know that the continuum description of  a fluid, say,  in terms of a set of dynamical variables like density $\rho$, velocity $\mathbf {v}$,  etc. has a  life of its own. At the same time, we also know that these dynamical variables and the
description have no validity at a fundamental level where the matter is discrete.
 But one can actually \textit{guess} the existence of microstructure without using 
 any experimental proof for the  molecular nature of the fluid, just from the fact that the 
 fluid or a gas  exhibits \textit{thermal phenomena} involving
 temperature and transfer of heat energy. If the fluid is treated as a continuum and is described
by  $\rho (t, \mathbf {x})$,  $ \mathbf {v}(t, \mathbf {x})$ etc., all the way down,   then it is \textit{not} possible to explain the thermal
phenomena in a natural manner. As first stressed by Boltzmann, the heat content of a fluid arises
due to random motion of discrete microscopic structures which \textit{must} exist in the fluid.
These new degrees of freedom --- which we now know are related to the actual
molecules --- make the fluid capable of storing energy internally and exchanging it
with surroundings. So, given an apparently continuum phenomenon which exhibits temperature, Boltzmann could infer the existence of underlying discrete degrees of freedom. 

\ha I agree. But what does it lead to in the present context?

\me The paradigm is: \textit{If you can heat it, it has microstructure!} And you can heat up spacetimes by collapsing matter or even by just accelerating \cite{daviesunruh}.
The horizons which arise in general relativity are endowed with temperatures \cite{hawking} which shows that, at least in this context,
some microscopic degrees of freedom are coming into play. 
So  a thermodynamic description that links the standard description of gravity with the statistical mechanics of --- as yet unknown ---
microscopic degrees of freedom must exist.
  It is in this sense that
I consider gravity to be emergent.

Boltzmann's insight about the thermal behaviour  has two other attractive features which are useful in our context. First, while the existence of the discrete degrees of freedom is vital in such an approach, the exact nature of the degrees of freedom is largely irrelevant. For example, whether we are dealing with argon molecules or helium molecules is largely irrelevant in the formulation of  gas laws and such differences can be taken care of in terms of a few well-chosen numbers (like, e.g.,  the specific heat). This suggests that such a description will have certain amount of robustness and independence as regards the precise nature of  microscopic degrees of freedom. 

Second, the entropy of the system arises due to our ignoring the microscopic degrees of freedom. Turning this around, one can expect the form of entropy functional to encode the essential aspects of microscopic degrees of freedom, even if we do not know what they are. If we can arrive at the appropriate form of entropy functional, in terms of some effective degrees of freedom, then we can expect it to provide the correct description.\footnote{Incidentally, this is why thermodynamics needed no modification due to either relativity or quantum theory. An equation like $TdS=dE+PdV$ will have universal applicability as long as effects of relativity or quantum theory are incorporated in the definition of $S(E,V)$ appropriately.}

\ha But most people working in quantum gravity will agree that there is some fundamental
microstructure to spacetime (``atoms of spacetime'') and the description of spacetime by metric is an approximate long distance description. So why are you making a big deal? I don't see anything novel here.

\me I will go farther than just saying there is microstructure and show you how to  actually use the thermodynamic concepts to provide an emergent description of gravity --- which no one else has attempted. If you think of
the full theory of quantum gravity as analogous to   statistical mechanics then  I will provide   the 
thermodynamic description of the same system. 

As you know, thermodynamics was developed
and used effectively decades before we knew anything about the molecular structure of matter or its statistical mechanics. Similarly, even without knowing the microstructure of spacetime or the full quantum theory of gravity, we can make lot of progress with the thermodynamic description of spacetime. 
The horizon thermodynamics, I will claim,
provides \cite{tworeviews} valuable insights about the nature of gravity totally independent of what ``the atoms of spacetime'' may be. It is somewhat like being able to describe or work with  gases or steam engines without knowing anything about the molecular structure of the gas or steam.

\section{Local Rindler observers and Entropy flow}\label{sec:lro}

\ha  All right. I hope all these will become clearer as we go along. Maybe I can suggest we plunge head-long into how \textit{you} would like to describe gravity. Then I can raise the issues as we proceed. 

\begin{figure}
	\begin{center}
	\includegraphics[scale=0.5]{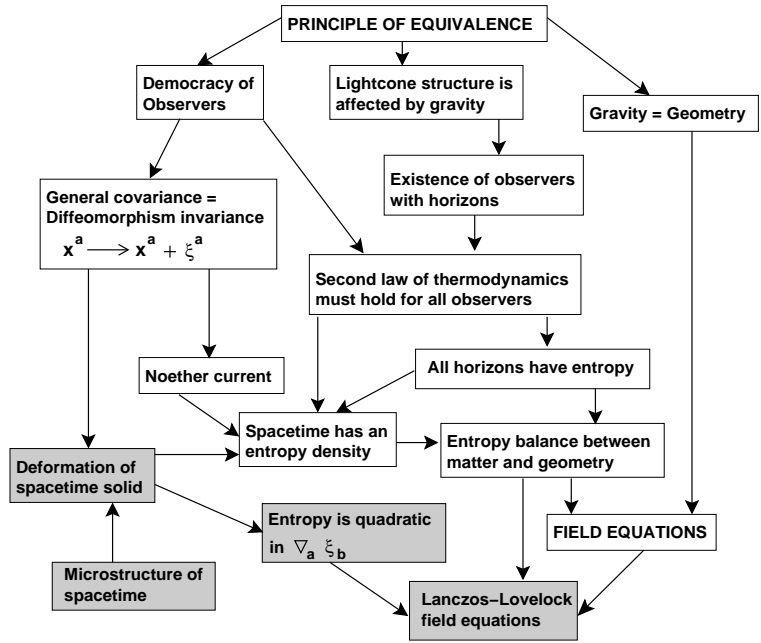}
	\end{center}

\caption{The broad picture }
\label{fig:synthesis}
\end{figure}

\me 
The overall structure of my approach is shown in Fig.~\ref{fig:synthesis}. As you can see, I begin with the principle of equivalence  which allows you to draw three key consequences. First, it tells you that --- in the long wavelength limit ---
gravity has \cite{mtw} a geometrical description in terms of the metric tensor $g_{ab}$ and the effect of gravity on matter can be understood by using the laws of  special relativity in the local inertial frames.
Second, by writing Maxwell's equations in curved spacetime using minimal coupling, say, I can convince myself that the light cone structure of the spacetime --- and hence the causal structure --- will, in general, be affected by the gravitational field.  

\ha Well, that is one possible way of interpreting principle of equivalence though people might have other views. But once you have told me what you are assuming, viz., ``gravity = geometry'' and ``light cones are affected by gravity'', we can proceed further. 
 
\me Yes. My aim here will be not  to nitpick over definitions but develop the physics in a consistent manner. In that spirit, I would draw one more conclusion from the fact that gravity can be described using a metric tensor.
In flat spacetime, we can choose a special coordinate system with the global metric being $\eta_{ab}$; so  if someone tells you that the metric is given by $g_{ab}(t,\mathbf{x})$ then
you can always attribute the part $(g_{ab}-\eta_{ab})$ to the choice of non-inertial coordinates. We cannot do this in a curved spacetime. So it no longer makes sense to ask ``how much of $g_{ab}$" is due to our using non-inertial coordinates and ``how much" is due to genuine gravity.  
Different observers in different states of motion might use different coordinates leading to different sets of ten functions for $g_{ab}(t,\mathbf{x})$. Because we have no absolute reference metric it follows that 
no coordinate system or  observer is special.  
The laws of physics should not select out any special class of observers. 

\ha This smacks of principle of general covariance but essentially you are arguing \cite{tpapoorva} for democracy of all observers, which I grant you. Given all the philosophical  controversies as to what `general covariance' means, I agree this is a safer procedure.
What next?

\me Given the fact that all observers are equal and that light cones are affected by gravity, it  follows that 
there will exist  observers  who do not have access to part of the spacetime  because of  the existence of horizons they perceive. This is  a direct consequence of the fact that metric determines the paths of light rays and hence the causal structure. The classic example is the Rindler horizon in flat spacetime which is as effective in blocking information with respect to an accelerated observer as the Schwarzschild horizon at $r=2M$ is for an observer at $r>2M$. 

\ha Not so fast; I have several problems here. First, the conventional view is that black hole horizons are ``real horizons'' while Rindler horizons are sort of fraudulent; you seem to club them together. Second, you seem to link horizons to observers rather than treat them as well defined, geometrical, causal features of a spacetime. 

\me You are quite right. I treat all horizons at equal footing and  claim 
that --- for \textit{my} purpose ---
all horizons are observer dependent. This is because, the key property of horizons which I am concentrating here is that it can block information. In that sense, 
the Rindler horizon does not block information for an inertial observer \textit{just as}  the  Schwarzschild horizon does \textit{not} block information for someone plunging into the black hole. 
So, for my purpose, there is no need to make artificial distinctions between a black hole  horizon and a Rindler horizon. 
The state of motion of the observer is crucial in deciding the physical effects of a horizon \textit{in all cases}.

It is, of course, true that one can give a geometric interpretation to, say, the black hole event horizon. I am not denying that. But that fact, as you will see, is quite irrelevant to the development of my approach. 

\ha I see that you not only demand democracy of observers but also democracy of horizons! You don't think, for example, that  black hole horizons are anything special.
 
 \me Yes. I do believe in the democracy of horizons, as you put it. The attempts to provide a quantum gravitational interpretation of black holes, their entropy etc. using \textit{very special} approaches which are incapable of handling other horizons ---  like the issues in de Sitter \cite{demohorizons}, let alone Rindler --- are interesting in a limited sort of way but may not get us anywhere ultimately.

\ha I have another problem.  You really haven't characterized what exactly you mean by a horizon
 for an observer. Of course, you cannot use any on-shell constructs since you are still developing your approach towards field equations. There are horizons and horizons in the literature --- event, apparent, causal ....
 
 \me I will try to make clear what I \textit{need} without again going into all sorts of definitions \cite{tp09papers}. Choose any event $\mathcal{P}$ and introduce a local inertial frame (LIF) around it with Riemann normal coordinates $X^a=(T,\mathbf {X})$ such that $\mathcal{P}$ has the coordinates $X^a=0$ in the LIF.  
Let $k^a$  be
 a future directed null vector at $\mathcal{P}$ and we align the coordinates of LIF
 such that it lies in the $X-T$ plane at $\mathcal{P}$. Next  transform from the LIF to a local Rindler frame (LRF) coordinates $x^a$ by accelerating along the X-axis with an acceleration $\kappa$ by the usual transformation. The metric near the origin now reduces to the form  
 \begin{eqnarray}
 ds^2 &=& -dT^2 + dX^2 + d\mathbf{x}_\perp^2\nonumber\\
 &=&-\kappa^2 x^2 dt^2 + dx^2 + d\mathbf{x}_\perp^2
= - 2 \kappa l \ dt^2 + \frac{dl^2}{2\kappa l}  + d\mathbf{x}_\perp^2
\label{surfrind}
\end{eqnarray} 
where $T=x \sinh (\kappa t);\  X= x \cosh (\kappa t);\ l=(1/2)\kappa x^2$
and ($t,x, \mathbf {x}_\perp$) or ($t,l, \mathbf {x}_\perp$)
are the coordinates of LRF (Both these forms are useful in our discussion).  
 Let $\xi^a$ be the approximate Killing vector corresponding to translation in the Rindler time such
that the vanishing of $\xi^a\xi_a \equiv -N^2$ characterizes the location of the 
local horizon $\mathcal{H}$ in LRF. As usual, we shall do all the computation 
on a timelike surface infinitesimally away from $\mathcal{H}$
with $N=$ constant, usually called a  ``stretched horizon''. (It 
can be defined more formally using  
the orbits of $\xi^a$ and the plane orthogonal to the acceleration vector $a^i= \xi^b \nabla_b\xi^i$.)
 Let the timelike unit normal to the stretched horizon
be  $r_a$. 

\begin{figure}
\centering
\scalebox{0.31}{\input{ray1.pstex_t}}\qquad\scalebox{0.31}{\input{ray2.pstex_t}}
\caption{The top frame illustrates schematically the light rays near an event
$\mathcal{P}$ in the $\bar t - \bar x$ plane of an arbitrary spacetime. The bottom frame
shows the same neighbourhood of $\mathcal{P}$ in the locally inertial frame at $\mathcal{P}$
in Riemann normal coordinates $(T,X)$. The light rays now become 45 degree lines and the trajectory of the local Rindler observer becomes a hyperbola very close to $T=\pm X$ lines
which act as a local horizon to the Rindler observer.}
\label{fig:ray1}
\end{figure}
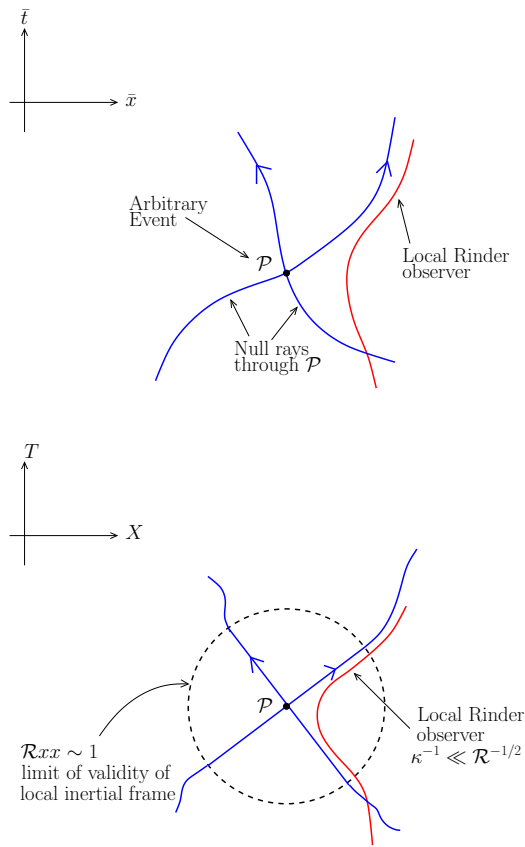

 This LRF (with  metric in \eq{surfrind}) and its local horizon $\mathcal{H}$ will exist within a region of size $L\ll\mathcal{R}^{-1/2}$ 
 (where $\mathcal{R}$ is a typical component of curvature tensor of the background spacetime) as long as $\kappa^{-1}\ll\mathcal{R}^{-1/2}$. This condition can always be satisfied by taking a sufficiently large $\kappa$. This procedure introduces a class of uniformly accelerated observers
 who will perceive the null surface $T=\pm X$ as the local Rindler horizon $\mathcal{H}$. This is shown in Fig. \ref{fig:ray1}.
 
 %%%%%%%%%%%%% Begin - Figure in pstex format %%%%%%%%%%%%%%
%%%% 

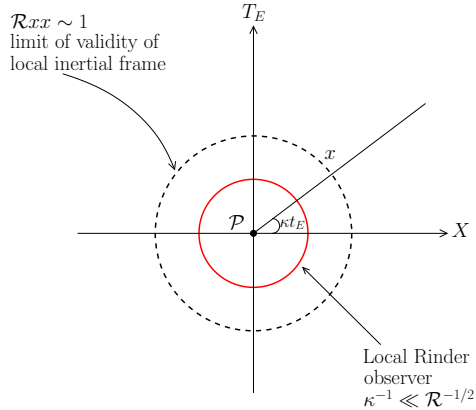
\begin{figure}
\begin{center}
\scalebox{0.31}{\input{ray3.pstex_t}}
\end{center}
\caption{The region around $\mathcal{P} $ shown in figure 1 is represented in the 
Euclidean sector obtained by analytically continuing to imaginary values of $T$ by $T_E = iT$. The horizons $T=\pm X$ collapse to the origin and the hyperbolic trajectory of the Rindler 
observer becomes a circle of radius $\kappa^{-1}$ around the origin. The Rindler coordinates
$(t,x)$ become --- on analytic continuation to $t_E = it$ --- the polar coordinates
$(r=x, \theta = \kappa t_E$) near the origin.}
\label{fig:ray3}
\end{figure}
%%%%
%%%%%%%%%%%   End - Figure in pstex format %%%%%%%%%

 \ha I am with you so far. Essentially you are using the fact that you have two length scales in the problem at any event. First is the length scale $\mathcal{R}^{-1/2}$ associated with the curvature components of the background metric over which you have no control; second is the length scale $\kappa^{-1}$ associated with the accelerated trajectory which you can choose as you please. So you can always ensure that  $\kappa^{-1}\ll\mathcal{R}^{-1/2}$. In fact, I can see this clearly in the  Euclidean sector in which the horizon maps to the origin (see Fig.\ref{fig:ray3}). The locally flat frame in the Euclidean sector will exist in a region of radius $\mathcal{R}^{-1/2}$ while the 
 trajectory of a uniformly accelerated observer will be a circle of radius $\kappa^{-1}$.
 You can always keep the latter inside the former. The metric in \eq{surfrind} is just the metric of the locally flat region in polar coordinates.
 
 \me Yes. In fact, I can choose a trajectory $x^i(\tau)$ such that its acceleration $a^j=u^i\nabla_i u^j$ (where $u^i$ is the time-like four velocity) satisfies the condition
 $a^ja_j =\kappa^2$. In a suitably chosen LIF this trajectory will reduce to the standard hyperbola of a uniformly accelerated observer. It is using these LRFs that I define my horizons around any event.
 Further the local temperature on the stretched horizon will be $\kappa/2\pi N$ so that 
$\beta_{\rm loc} = \beta N$ with $\beta \equiv \kappa/2\pi$.

 \ha Ha! The classical GR is fine but your `horizon' is just a patch of null surface.  Can you actually prove that local Rindler observers will perceive a temperature proportional to acceleration ? The usual proofs of Unruh effect are \cite{daviesunruh} horribly global.
 
 \me Recall that everything we do is in a local region with $\kappa^{-1}\ll\mathcal{R}^{-1/2}$. Now if you have an accelerated detector with time-dependent, variable acceleration, say, then you will reproduce the standard Unruh effect \textit{approximately} to the necessary order of accuracy. This should be intuitively obvious but can be demonstrated \cite{dawood-paddy-varg}.
 Of course in the Euclidean sector the Rindler observer's trajectory is a circle of radius $\kappa^{-1}$ which can be made arbitrarily close to the origin. Suppose the observer's trajectory has the usual form $X=\kappa^{-1}\cosh \kappa t; T=\kappa^{-1}\sinh \kappa t$ which is maintained for a time interval  $T\approx 2\pi/\kappa$. Then, the trajectory will complete a full circle \textit{in the Euclidean sector} irrespective of what happens later! When we work in the limit of $\kappa\to\infty$, this becomes arbitrarily local in both space \textit{and} time \cite{loctemp}. I am sure all these can be made more rigorous but this is the essential reason behind the local ideas working.

 I also want to stress that once I finally reach my goal (of deriving the gravitational field equations from an entropy principle in Section \ref{sec:eqnnewvar}) all these become irrelevant; they are essentially part of `motivation'. So your possible misgivings regarding some of these details will not affect the final result.
 
 \section{Thermodynamic re-interpretation of the field equations}\label{sec:reinterpret}
 
 \ha OK, so you have local Rindler observers crawling all over the spacetime with their local horizons. What next ?

\me It is now  
 easy to see that all horizons must have entropy vis-a-vis the observers who perceive the horizons. If they do not, an observer can pour some hot tea with entropy across the horizon \textit{\`a la Wheeler} \cite{wheeler}, thereby violating the second law of thermodynamics in the region accessible to her and her friends who perceive the horizon $\mathcal{H}$.  
Such a violation
  of second law of thermodynamics can be avoided only if we  \textit{demand} that
horizon should have an entropy which should increase when energy flows across it.
If energy $dE$ flows across a hot horizon with temperature $T$ then $dE/T=dS$ should be the change in  the entropy of the horizon.
We therefore conclude that all null surfaces which could locally act as one-way membranes should have an (observer dependent) entropy associated with them.  
 
 \ha Hold on.  I understand from the reference you cite  \cite{wheeler} that such a thought experiment might have had something to do with the initial realization of a black hole entropy which is proportional to the area  \cite{bekenstein}. But I am not sure how to interpret it precisely.  For one thing, matter  disappears into the horizon
 only after infinite time as perceived by the outside observer, even when you try to pour real tea in to real black hole. So what is all this talk about ``loss'' of entropy? 
 
 \me I don't think this is a real objection though one often comes across this confusion.
 Note that, by the same argument, no black hole can ever form in finite time anywhere in the universe and we should not be talking about any black hole physics. I believe this issue is well settled in Chapter  33  of Ref.~\cite{mtw}. I recommend you read it!
 
 If you really push me hard, I can wiggle out with the following argument. It does not 
 take much time (certainly not infinite time!) for a cup of tea to reach a
 radial distance a few Planck lengths away from the horizon $r=2M$. We have considerable evidence of very different nature to suggest Planck length acts as lower bound to the length scales that can be operationally defined and that no measurements can be ultra sharp at Planck scales \cite{tplimitations}. So one cannot really talk about the location of the event horizon ignoring fluctuations of this order. So, from the point of view of sensible physics, I only need to get the cup of tea up to $r=2M +L_P$ to talk about entropy loss. 
 
\ha You also seem to have quietly made entropy an observer dependent quantity. \textit{This is pretty drastic} and I need to understand it. Suppose, in a region around an event $\mathcal{P}$, there is some matter which is producing the curvature.  I would have normally thought that this matter --- say some amount of hot fluid --- has certain amount of entropy which is independent of who is measuring it. But you are claiming that an inertial observer and a Rindler observer will attribute different amounts of entropy to this matter. Is that correct?
 
  \me That's correct and may be I should write a paper explaining this \cite{tpinprogress} but 
  this really need not come as a surprise (also see ref.\cite{marolf}). We know that an inertial observer will attribute
  zero temperature and zero entropy to the inertial vacuum. But a Rindler observer will attribute a finite temperature and non-zero (formally divergent) entropy to the same vacuum state. So entropy is indeed an observer dependent concept. When you do quantum field theory in curved spacetime, it is 
  not only that particles become an observer dependent notion so do the  temperature and entropy. This notion can be made
  more precise as follows: 
  
  Consider an excited state of a quantum field with energy  $\delta E$ above the ground state in an inertial spacetime. When you integrate out the unobservable modes for the Rindler observer, you will get a density matrix $\rho_1$ for this state
  and the corresponding entropy will be $S_1 = - {\rm Tr}\ (\rho_1 \ln \rho_1)$.
  The inertial vacuum state has the density matrix $\rho_0$ and the entropy $S_0 = - {\rm Tr}\ (\rho_0 \ln \rho_0)$. The difference $\delta S = S_1 - S_0$ is finite and 
  represents the entropy attributed to this state by the Rindler observer. (This is
  finite though $S_1$ and $S_0$ can be divergent.) In the limit of $\kappa \to \infty$,
  in which we are working, we can actually compute it and show that 
  \begin{equation}
\delta S = \beta \delta E = \frac{2\pi}{\kappa} \delta E
\label{delS}
\end{equation}
To see this, note that if we write $\rho_1=\rho_0 + \delta \rho$, then in the limit of
$\kappa \to \infty$ we can concentrate on states for which $\delta \rho/\rho_0\ll 1$.
Then we have
\begin{eqnarray}
-\delta S &=& {\rm Tr}\ (\rho_1 \ln \rho_1) - {\rm Tr}\ (\rho_0 \ln \rho_0) 
\simeq {\rm Tr}\ (\delta \rho \ln \rho_0)\nonumber\\
&=& {\rm Tr}\ (\delta \rho (-\beta H_R)) = - \beta {\rm Tr}\ \left((\rho_1 -\rho_0)H_R\right) \equiv -\beta \delta E
\end{eqnarray} 
where we have used the facts Tr $\delta \rho \approx 0$ and
$\rho_0 =Z^{-1}\exp(-\beta H_R)$ where $H_R$ is the Hamiltonian for the system in the 
Rindler frame. The last line defines the $\delta E$ in terms of the difference in 
the expectation values of the Hamiltonian in the two states. 
(There are some subtleties in this derivation, especially regarding the assumption
$\delta \rho/\rho_0\ll 1$, but I will not get into it here \cite{tpinprogress}.) 
This is the amount of entropy a Rindler observer would claim she has lost when the matter
disappears into the horizon.

\ha That is very curious. I would have thought that the expression for entropy of matter
should consist of its energy $\delta E$ and \textit{its own} temperature $T_{\rm matter}$
rather than the horizon temperature. It looks like that the matter somehow equilibrates
to the horizon temperature so that $\delta S = \delta E/T_{\rm horizon}$ gives the 
relevant entropy.

\me Yes. This can be explicitly proved, for example, for the one particle state \cite{sanved} and
 here is a possible interpretation.
You should think of horizon as a system with some internal degrees of freedom and temperature $T$ \textit{as far as Rindler observer is concerned}. So when you add an energy $\delta E$ to it, the entropy change is $\delta S = (\delta E/ T)$.
All these are not \textit{new} mysteries but only the manifestation of the old mystery, viz., a Rindler observer attributes a non-zero temperature to inertial vacuum. This temperature influences every other thermodynamic variable. 
I will come back to this point later on because it is quite important.

\ha OK. Let us proceed. I also see where your insistence of democracy of observers comes in. You want to demand
that the local Rindler observer has a right to expect the standard laws of physics to hold as much as any other observer, horizons notwithstanding.

\me I am glad you brought this up.  This was first pointed out in Ref.~\cite{tpapoorva}
in which we  assert that {\it all observers have a right to describe physics using an effective theory based only on the variables she can access.}
In the study of particle physics models, this concept forms the cornerstone
of  the  renormalization group theory. 
 To describe particle interactions  at 10 GeV in the lab, we usually do not need to know what happens at $10^{14}$ GeV in  theories which have predictive power.
 In the absence of such a principle, very high energy phenomena (which are unknown
 from direct experiments in the lab) will affect the low energy phenomena which
 we are attempting to study. 
 
 In the context of a theory involving a  nontrivial metric of spacetime, we need a similar principle
 to handle the fact that different observers will have access to different regions of a  general spacetime.
  If a class of observers perceive a horizon, they should still be able to do physics using only the variables accessible to them without having to know what happens on the other side of the horizon.\index{surface term!horizon entropy}

  This, in turn, implies that there should exist a mechanism which will encode
  the information in the region $\mathcal{V}$ which is inaccessible to a particular observer at the boundary $\partial \mathcal{V}$ of that region. 
  Keep this in mind because I will show you later where this fits in with the holographic nature of action functionals.
  
\ha Fine, we will get back to it. To get on with the story, you need to formulate some kind of entropy balance when matter flows across a local horizon. How do you propose to do it ?

\me  Around any event in any spacetime we now have a local inertial frame and --- by boosting along one of the axes with an acceleration $\kappa$ --- we have introduced a local Rindler observer who perceives a horizon with temperature proportional to $\kappa$. 
She will attribute a loss of entropy $\delta S = (2\pi/\kappa) \delta E$  when matter with an amount of energy $\delta E$ gets close to the horizon
 (within a few Planck lengths, say). If $\xi^a$ is the  approximate,  Killing vector corresponding to translations in Rindler time,  the appropriate energy-momentum  density is $T^a_b\xi^b$. (It is the integral of $T^a_b\xi^b d\Sigma_a$ that gives the Rindler Hamiltonian $H_R$, which leads to evolution in Rindler time $t$ and appears in the thermal density matrix $\rho=\exp-\beta H_R$.) The energy flux through a
 patch of stretched horizon with normal $r_a$ will be $T_{ab}\xi^ar^b$ and the associated entropy flux will be $\beta_{loc} T_{ab}\xi^ar^b$ where $\beta_{loc}^{-1}=\beta^{-1}/N$ is the local temperature with $N$ being the standard lapse function giving the redshift factor. (In conformity with \eq{delS}, I am using the horizon temperature and not the matter temperature).
 This entropy flux manifests as  the entropy change of the locally perceived  horizon.  For all these to hold locally at every event there \textit{must exist a spacetime entropy current}  $\beta_{loc}J^a$, built out of metric and its derivatives, such that
 $\beta_{loc}(r_aJ^a)$  gives the corresponding  gravitational entropy flux.  So we expect the relation
  \begin{equation}
\beta_{loc} r_a J^a =\beta_{loc} T^{a b} r_a \xi_b
\label{crucial1}
\end{equation}  
 to hold at 
 all events with some $J^a$, once we introduce a local Killing vector $\xi^a$ and a local temperature giving $\beta_{loc}$. Further $J^a$ must be conserved since we do not expect irreversible entropy production in the spacetime. 

 \ha This sounds strange!. Why should there exists a conserved current $J^a$, built from geometrical variables, at every event
 in some arbitrary spacetime, which will conveniently give you the entropy balance you
 require?
 
 \me It is actually not all that  strange! Remember that we got into all these because of the democracy of the observers which, in turn, implies general covariance.  
 The mathematical content of general covariance is captured by the diffeomorphism
 invariance of whatever theory which is going to ultimately determine the 
 dynamics of the spacetime.  Because  the  diffeomorphism invariance of the theory forced us to treat all observers on equal footing,   the diffeomorphism invariance must also provide us with the conserved current $J^a$.  And indeed it does, in the form of the Noether current \cite{tp09papers}. Let me explain.

 Consider a theory of gravity, obtained from a generally covariant action principle
involving a gravitational Lagrangian $L(R^{a}_{bcd}, g^{ab})$ which is a scalar made
from metric and curvature tensor. The total Lagrangian is the sum of $L$
and the matter Lagrangian $L_m$. The variation of the gravitational Lagrangian
density generically leads to a surface term and hence can be expressed in the
form,
\begin{equation}
\delta(L\sqrt{-g }) =\sqrt{-g }\left( E_{ab} \delta g^{ab} + \nabla_{a}\delta v^a\right). \label{variationL} 
\end{equation}
Under suitable boundary conditions  the theory will lead to the field equation $2 E_{ab} = T_{ab}$ where $E_{ab}$ is given by \eq{genEab} and  $T_{ab}$ is defined through the usual relation $(1/2)T_{ab}\sqrt{-g}=-(\delta A_{m}/\delta g^{ab})$. We also know that, for any Lagrangian $L$, the functional
derivative $E_{ab}$ satisfies the generalized off-shell Bianchi identity:
$
\nabla_a E^{ab} = 0. 
$

Consider now the variations in $\delta g_{ab}$ which arise through the diffeomorphism $x^a \rightarrow x^a + \xi^a$. In this case, $\delta (L\sqrt{-g} ) = -\sqrt{-g} \nabla_a (L \xi^a)$, with $\delta g^{ab} = (\nabla^a \xi^b + \nabla^b \xi^a)$. Substituting these in \eq{variationL} and using $
\nabla_a E^{ab} = 0 
$, we obtain the
conservation law $\nabla_a J^a = 0$, for the current,
\begin{equation}
J^a \equiv \left(2E^{ab} \xi_b + L\xi^a + \delta_{\xi}v^a   \right) 
\label{current}
\end{equation}
where $\delta_{\xi}v^a$ represents the boundary term which arises for the specific variation
of the metric in the form $ \delta g^{ab} = ( \nabla^a \xi^b + \nabla^b \xi^a$). 
It is also convenient to introduce the antisymmetric tensor $J^{ab}$ by $J^a = \nabla_b J^{ab}$.
Using the known expression for $\delta_{\xi}v^a $ in \eq{current}, it is possible to write an explicit expression for the current $J^a$ for any diffeomorphism invariant theory. For the general
class of theories we are considering, the $J^{ab}$ and $J^a$ can be expressed \cite{mohut} in the form
\begin{equation}
J^{ab} = 2 P^{abcd} \nabla_c \xi_d - 4 \xi_d \left(\nabla_c P^{abcd}\right)
\label{noedef}
\end{equation} 
\begin{equation}
J^a = -2 \nabla_b \left (P^{adbc} + P^{acbd} \right ) \nabla_c \xi_d + 2 P^{abcd} \nabla_b \nabla_c \xi_d - 4 \xi_d \nabla_b \nabla_c P^{abcd} 
\end{equation} 
where $P_{abcd}\equiv (\partial L/\partial R^{abcd})$. These expressions simplify significantly at any event $\mathcal{P}$ where  $\xi^a$ behaves like an (approximate) Killing vector and satisfies the conditions
\begin{equation}
 \nabla_{( a} \xi_{b)} = 0;\quad  \nabla_a \nabla_b \xi_c = R_{c b a d} \xi^d
 \label{cond1}
\end{equation}
(which a true Killing vector will satisfy everywhere).
Then one can easily prove that
$
\delta_{\xi}v^a=0
$
at the event $\mathcal{P}$;  
the expression for Noether current simplifies considerably and is given by
\begin{equation}
 J^a \equiv \left( 2E^{ab} \xi_b + L\xi^a \right). 
 \label{current1}
\end{equation}

 \ha OK. So you now have a conserved current  $J^a$ and entropy current of matter. What do you do now ?
 
 \me Recall that I argued, on very general grounds,  that the relation in \eq{crucial1} \textit{must} hold at all events. 
 Remarkably enough, \textit{the gravitational field equations of any diffeomorphism invariant theory
 implies that this relation does hold!} 
   To see this, let us now consider the form of $J^{a}(x)$ at any event ${\cal P}$ around which we have introduced  the notion of a local Rindler horizon
 with  $\xi^a$ being  the approximate Killing vector associated with the Rindler time translation invariance that  satisfies two conditions in \eq{cond1} at ${\cal P}$. 
Let $r_a$ be the spacelike unit normal to the stretched horizon $\Sigma$, pointing in the direction of increasing $ N$. We know that as $ N\to0$ and the stretched horizon approaches the local horizon and $ N r^i$ approaches $\xi^i$.

With this background, we compute $J^a$ for the $\xi^a$ introduced above in the neighborhood of $\mathcal{P}$. Since it is an approximate  Killing vector, satisfying \eq{cond1} it follows that $\delta_\xi v=0$ giving the current to be
$
J^a = \left(2 E^{a b} \xi_b + L \xi^a \right) $.
 The product $r_a J^a$ for the vector $r^a$, which satisfies $\xi^ar_a=0$ on the stretched horizon,
becomes quite simple:
$
r_a J^a = 2 E^{a b} r_a \xi_b $.
This equation is valid around the local patch in which $\xi^a$ is the approximate Killing vector. The quantity $ \beta_{loc} r_a J^a$ (in this limit) is what we interpret as the local entropy flux density. On using the field equations $2E_{ab}=T_{ab}$, we  immediately get
\begin{equation}
\beta_{loc} r_a J^a =2 E^{a b} r_a \xi_b=\beta_{loc} T^{a b} r_a \xi_b
\label{final}
\end{equation}
which is exactly \eq{crucial1}. This tells you that the validity of field equations in any diffeomorphism invariant theory has a local, thermodynamic, interpret ion.
In the limit of $ N\to0$, this gives a \textit{finite} result, $\beta \xi_a J^a =\beta T^{a b} \xi_a \xi_b$ as it should.
Further, in this limit, $\xi^i$  goes to $\kappa \lambda k^i$ where $\lambda $ is the affine parameter associated with the null vector $k^a$ we started with  and all the reference to LRF goes away.  
 It is clear that the properties of LRF are relevant conceptually to define the intermediate notions (local Killing vector, horizon temperature ....) but the essential result is independent of these notions. Just as we introduce local inertial frame to decide how gravity couples to matter, we use local Rindler frames to interpret the physical content of the field equations.

 \ha That is cute! I also see why you can afford to be a bit cavalier about the LRF etc; ultimately, your interpretation is local at each event.
 The Noether current you use, of course, is the same that appears in the definition of Wald entropy \cite{wald}. But in the latter, it is used in an integral form while your approach seems to be completely local. 

\me This is true and I think the local approach is crucial for proper interpretation.
Integrals over surfaces would require all sort of special assumptions for everything to work out in an arbitrary spacetime. This is why I work in a local region around an arbitrary event with LIF, LRF etc. with L in everything. 
Also note that  the
original definition of Wald entropy is an on-shell construct and requires you to evaluate an integral on a solution. The Noether current itself is an off-shell construct and that is what I need. 

Incidentally, the Noether current relation can also be used to provide an alternative interpretation of the entropy balance along the following lines. 
 A local Rindler observer, moving along the orbits of the Killing vector field $\xi^a$ with four velocity $u^a = \xi^a/N$, will associate  an energy
  $\delta E =u^a(T_{ab}\xi^b) dV_{\rm prop}$ with a proper volume $dV_{\rm prop}$.
  If this energy gets transfered across the horizon, the corresponding entropy transfer will be $\delta S_{\rm matter} = \beta_{\rm loc}\delta E$ where $\beta_{\rm loc} = \beta N = 
(2\pi/\kappa)N$ is the local  (redshifted) temperature of the horizon
 and $N$ is the lapse function. 
 Since $\beta_{\rm loc} u^a = (\beta N)(\xi^a/N) = \beta \xi^a$, we find that 
\begin{equation}
 \delta S_{\rm matter} = \beta \xi^a \xi^b T_{ab}\ dV_{\rm prop}
 \label{defsmatter}
\end{equation} 
As for gravitational entropy, since $J^0$ is the Noether charge density, $\delta S = \beta_{\rm loc} u_a J^a dV_{\rm prop}$ can be interpreted
as the entropy associated with a volume $dV_{\rm prop}$ as measured
by an observer with four-velocity $u^a$. For observers moving on the orbits of the Killing vector $\xi^a$
with $u^a = \xi^a/N$  we get
\begin{equation}
\delta S_{\rm grav} = \beta N u_a J^a dV_{\rm prop}  =\beta  [\xi_j \xi_a T^{aj} +  L  (\xi_j \xi^j)]\, dV_{\rm prop}
\end{equation} 
As one approaches the horizon, $\xi^a\xi_a\to 0$ making the second term vanish and we 
get 
\begin{equation}
\delta S_{\rm grav} = \beta  (\xi_j \xi_a T^{aj}) \, dV_{\rm prop} = \delta S_m
\end{equation}
In the same limit $\xi^j$ will become proportional to the original null vector $k^j$ we started with.   
 So this 
equation can be again thought of as an entropy balance condition. 

\ha So instead of thinking of field equations of gravity as $2E_{ab}=T_{ab}$, you want us to think of them as 
\begin{equation}
 [2E^{ab} - T^{ab}]k_ak_b =0
 \label{myeqn}
\end{equation}
for all null vectors $k^a$.
This is equivalent to $2E^{ab} - T^{ab}=\lambda g^{ab}$ with some constant $\lambda$.
(I see that the constancy of $\lambda$ follows from the conditions $\nabla_aE^{ab}=0,\ \nabla_a T^{ab}=0$.) Interpreting $2E_{ab}k^ak^b$ as some kind of gravitational entropy density and $T_{ab}k^ak^b$ as matter entropy in the local Rindler frame, you are
are providing a purely thermodynamical interpretation of the 
 field equations of any diffeomorphism invariant theory of gravity. Right ?
 
 \me Yes. But note
  that  
 \eq{myeqn} is \textit{not} quite the same as the standard equation $2E^{ab}=T^{ab}$ because \eq{myeqn} has an extra symmetry
 which standard gravitational field equations do not have: This equation is invariant
 under the shift $T^{ab}\to T^{ab}+\mu g^{ab}$ with some constant $\mu$. (This symmetry has important implications   for cosmological constant problem which we will discuss later.) While the properties of LRF are relevant conceptually to define the intermediate notions (local Killing vector, horizon temperature ....),  the essential result is independent of these notions. 
 
 \ha Fine. I like the fact that
 \textit{just as we introduce local inertial frames to decide how gravity couples to matter, we use local Rindler frames to interpret the physical content of the field equations.} 
 But you only needed the part of $J^a$ given by $2E^a_b\xi^b$ for your analysis, right ? The other two terms in \eq{current} are not needed at all. So may be you don't have to use all of Noether current.

\me This is quite true. In fact one can give $2E^a_b\xi^b$ an interesting interpretation. Suppose there are some microscopic degrees of freedom in spacetime, just as there are atoms in a solid. If you make the solid undergo an elastic deformation $x^\alpha\to x^\alpha+\xi^\alpha(x)$, the physics can be formulated in terms of the displacement field $ \xi^\alpha(x)$ and one can ask how thermodynamic potentials like entropy change under such displacement. Similarly, in the case of spacetime, we should think of
\begin{equation}
\delta S_{grav}=\beta_{loc}(2E^a_b)u_a\delta x^b
\end{equation} 
 as the change in the gravitational entropy under the `deformation' of the spacetime $x^a\to x^a+\delta x^a$ as measured by the Rindler observer with velocity $u^a$. One can show that this interpretation is consistent with all that we know about horizon thermodynamics. So the left hand side of gravitational field equation $(2E^a_b)$ actually gives the response of the spacetime entropy to the deformations. 

 \ha It certainly matches with the previous results. Since $ \beta_{loc} u_a=\beta\xi_a$, you will get the entropy density to be proportional to $2E_{ab}k^ak^b$ on the horizon. Does it make sense ?
 
 \me As I will show you soon, it makes lot of sense!

 \ha  But can't you now reverse the argument and claim that you can derive the field equations of the theory from the purely thermodynamic point of view of the entropy balance? 
 
 \me That would be lovely and very tempting but I don't think so. 
 Such a 
`reverse engineering' faces   some conceptual hurdles; the mathematics will go through trivially but not the logic \cite{tp09papers}. Let me clarify the issues involved. 

 The key point is the following:  If we have a justification for interpreting the expression $\beta_{loc}(r_aJ^a)$ as entropy current, {\it independent of the field equations}, then --- {\it and only then} --- can we invert the logic and obtain the field equations from the thermodynamic identity. However, in the absence of field equations $J^a$ is just a Noether current. It can be interpreted as entropy current {\it if and only if} field equations are assumed to hold; it is in this on-shell context that Wald \cite{wald} showed that it is  entropy. So we have no {\it independent} justification for demanding $\beta_{loc} r_aJ^a$ should be equal to matter entropy flux. Until we come up with such a justification --- without using field equations --- we can prove that ``field equations imply local entropy balance at local horizons" but not ``local entropy balance at local horizons imply field equations". The issue at stake is not mathematics but logic. 
 As a simple example, consider the Noether current in Einstein's theory for a Killing vector $\xi^a$, which is proportional to $R^a_b \xi^b$. No one would have thought of this expression as entropy density independent of field equations. It is only by studying physical processes involving black holes, say, \textit{and using field equations} that one can give such a meaning.

 \ha OK. I have one more worry. At this stage, you have not chosen any specific theory of gravity at all, right? So this thermodynamic entropy balance seems to be very general and some people might even say it is \textit{too general}. What is your take on this?
  
  \me It is true that at this stage I have not specified
  what kind of theory of gravity we are dealing with.
The field equation --- whatever the theory may be, as long as it obeys principle of equivalence and diffeomorphism invariance --- always has an interpretation in terms of local entropy balance (The idea also works when $L_{\rm grav}$ depends on the derivatives of the curvature tensor but I will not discuss this case, for the sake of simplicity.) 
 \textit{Different theories of gravity are characterized by different forms of entropy density just as different physical systems are characterized by different forms of entropy functionals.}
I think this is completely  in harmony with the thermodynamic spirit. Thermodynamics applies to any system; if you want to describe a \textit{particular} system, you need to specify its entropy functional or some other thermodynamic potential. So what the development so far is telling us is that we need to put in some more extra physical input into the theory to find the field equations describing the theory.

\section{Field equations from a new variational principle}\label{sec:eqnnewvar}

\ha Fine. The above results imply that the field equations arising from any generally covariant action can be given a thermodynamic interpretation; that is, you assumed the validity of the  field equations and derived the local entropy balance. Your real aim, however, is to obtain the field equations from a dynamical principle rather than \textit{assume} the field equations. How do you propose to do that ?

\me To begin with, I want to paraphrase the above results in a slightly different manner which is probably more useful for the task we want to undertake.  
 
Note that, instead of dropping matter across the horizon, I could have equally well
  considered a  virtual, infinitesimal (Planck scale), displacement of the $\mathcal{H}$ normal to itself
engulfing some matter. We only need to consider infinitesimal displacements because the entropy of the matter is not `lost' until it crosses the horizon; that is, until when the matter is at an infinitesimal distance (a few Planck lengths) from the horizon. All the relevant physical processes take place at a region very close to the horizon  and hence  an infinitesimal displacement of $\mathcal{H}$ normal to itself will engulf 
some matter. 
 Some entropy will be again lost to the  outside observers unless  displacing a piece of local Rindler horizon  costs some entropy. 
 
 So we expect the entropy balance condition  derived earlier to ensure this and indeed it does.  
An infinitesimal displacement of a local patch of the stretched horizon in the direction of $r_a$, by an infinitesimal proper distance $\epsilon$, will change the proper volume by $dV_{prop}=\epsilon\sqrt{\sigma}d^{D-2}x$ where $\sigma_{ab}$ is the metric in the transverse space.
 The flux of energy through the surface will be  $T^a_b \xi^b r_a$ and the corresponding  entropy flux
 can be obtained by multiplying the energy flux by $\beta_{\rm loc}$.  Hence
 the `loss' of matter entropy to the outside observer when the virtual displacement of the horizon swallows some hot tea is 
$\delta S_m=\beta_{\rm loc}\delta E=\beta_{\rm loc} T^{aj}\xi_a r_j dV_{prop}$. 
To find the change in the gravitational entropy, we again use the Noether current $J^a$ corresponding
to the local Killing vector $\xi^a$. 
Multiplying by $r^a$ and 
 $\beta_{\rm loc} = \beta N$, we get
\begin{equation}
\beta_{\rm loc} r_a J^a  = \beta_{\rm loc}\xi_a r_a T^{ab}  + \beta N ( r_a \xi^a) L
\end{equation}  
As the stretched horizon approaches the true horizon, we know that  $N r^a \to \xi^a$
 and $\beta \xi^a \xi_a L \to 0$ making the last term vanish. So
\begin{equation}
\delta S_{\rm grav} \equiv  \beta \xi_a J^a dV_{prop} = \beta T^{aj}\xi_a \xi_j dV_{prop}
=\delta S_m
\end{equation} 
showing again the validity of local entropy balance.

\ha It appears to me that this is similar to switching from a passive point of view to an active point of view. Instead of letting a cup of tea fall into the horizon, you are making a virtual displacement of the horizon surface to engulf the tea which is infinitesimally close to the horizon. But in the process, you have introduced the notion of virtual displacement of horizons and for the theory to be consistent, this displacement
of these surface degrees of freedom should cost you some entropy. Right?

\me Yes. If 
gravity is an emergent, long wavelength, phenomenon like elasticity  then the diffeomorphism $x^a\to x^a+\xi^a$ is analogous to  the elastic deformations of the ``spacetime solid"
\cite{elasticgravity}. It then makes sense to demand that the entropy density should be a functional of $\xi^a$ and their derivatives $\nabla_b \xi^a$. By constraining 
the functional form of this entropy density, we can choose the field equations of gravity.
Recall that thermodynamics relies entirely on the form of the entropy functional to make predictions. If we constrain the form of the entropy, we constrain the theory.

So the next step is  to assume a suitable form of entropy functional for gravity  $S_{grav}$ in terms of the normal to the null surface. Then it seems
 natural to demand that  the dynamics should follow from the extremum prescription
$\delta[S_{grav}+S_{matter}]=0$ for \textit{all null surfaces in the spacetime} where $S_{matter}$ is the matter entropy. 

\ha What do we take for $S_{grav}$ and $S_{matter}$ ?

\me The form of $S_{matter}$ is easy to ascertain from the previous discussion.
If $T_{ab}$ is the matter energy-momentum tensor in a general $D(\ge 4)$ dimensional spacetime then an expression for matter entropy \textit{relevant for our purpose} can be taken to be 
\begin{equation}
S_{\rm matt}=\int_\Cal{V}{d^Dx\sqrt{-g}}
      T_{ab}n^an^b
      \label{Smatt}
\end{equation} 
where $n^a$ is a null vector field.  
From our \eq{defsmatter} we see that the entropy density associated with proper 3-volume is $\beta(T_{ab}\xi^a\xi^b)dV_{prop}$ where --- on the horizon --- the vector $\xi^a$ becomes proportional to a null vector $n^a$. 
If we now use the Rindler coordinates in \eq{surfrind} in which $\sqrt{-g}=1$ and 
interpret the factor $\beta$ as arising from an integration of $dt$ in the range $(0,\beta)$ we find that the entropy density associated with a proper four volume is $(T_{ab}n^an^b)$. This suggests treating \eq{Smatt} as the matter entropy. 
 For example, if $T_{ab}$ is due to an ideal fluid at rest in  the LIF then $T_{ab}n^an^b$ will contribute $(\rho+P)$, which --- by Gibbs-Duhem relation --- is just $T_{local}s$ where $s$ is the entropy density and $T_{local}^{-1}=\beta N$ is the properly redshifted temperature with $\beta=2\pi/\kappa$ being the periodicity of the Euclidean time coordinate. Then
 \begin{eqnarray}
\int dS&=&\int\sqrt{h}d^3x\ s=\int\sqrt{h}d^3x\beta_{\rm loc}(\rho+P)=\int\sqrt{h}Nd^3x \beta (\rho+P)\nonumber\\
&=&\int_0^\beta dt \int d^3x\g T^{ab}n_an_b
\label{intlim}
\end{eqnarray} 
which matches with \eq{Smatt} in the appropriate limit.

\ha Well, that may be all right for an ideal fluid. But for a general source, like say  the electromagnetic field (which will act as a source in \rn\ metric) I don't even know how to define entropy.
But I am willing to accept \eq{Smatt} as a definition.

\me Actually, it is better than that. We \textit{do} have the notion of \textit{energy} flux across a surface with normal $r^a$ being $T_{ab}\xi^br^a$ which holds for \textit{any} source $T^{ab}$.
 Given some energy flux $\delta E$ in the Rindler frame, there is an associated entropy flux loss $\delta S=\beta_{hor}\delta E$ as given by \eq{delS}. You may think that an ordered field  has no temperature or entropy but a Rindler observer will say something different. For any state, she will have a corresponding density matrix $\rho$ and an entropy $-Tr(\rho\ln\rho)$; after all, she will attribute entropy even to vacuum state. It is \textit{this} entropy which is given by \eq{defsmatter} and \eq{Smatt}. 

\ha Interesting. There is also this time integration which you limit to the range $(0,\beta)$ in \eq{intlim}. This is fine in Euclidean sector and may be you can rotate back to Lorentzian sector but it makes me a little uncomfortable. 

\me Well, I again have to invoke local nature of the argument which, as we discussed, is obvious in Euclidean sector but the concept of causality, loss of information etc are obvious in the Lorentzian sector in which I have light cones and null surfaces. So I do have to switch back and forth. May be there is a better way of formulating this which I have not  yet figured out; but for our purpose you can even think of all integrals being done in Euclidean sector --- if you are happier with that. 
 
\ha Fine. What about $S_{grav}$ ?

\me For this, I will first describe the simplest possible choice and will then consider a more general expression. The simplest choice is to  postulate $S_{grav}$ to be a quadratic expression \cite{aseementropy} in the derivatives of the normal:  
 \begin{equation}
S_{grav}= - 4\int_\Cal{V}{d^Dx\sqrt{-g}}
    P_{ab}^{\ph{a}\ph{b}cd} \D_cn^a\D_dn^b 
    \label{Sgrav}
\end{equation}  
where the explicit form of $P_{ab}^{\ph{a}\ph{b}cd}$ is ascertained below. The expression for the total entropy,  now becomes:
\begin{equation}
S[n^a]=-\int_\Cal{V}{d^Dx\sqrt{-g}}
    \left(4P_{ab}^{\ph{a}\ph{b}cd} \D_cn^a\D_dn^b - 
    T_{ab}n^an^b\right) \,,
\label{ent-func-2}
\end{equation}
If you want, you can forget everything we said so far and start with this expression as defining our theory!

\ha I suppose this is your variational principle and you will now extremise $S$ with respect to $n_a$. 

\me Yes, but
 I want to first explain the crucial conceptual difference 
between the extremum principle introduced here and the conventional one. Usually, given a set of dynamical variables $n_a$ and a functional $S[n_a]$, the extremum principle will give a set of equations for the dynamical variable $n_a$. Here the situation is completely different. We expect the variational principle to hold for   \textit{all} null vectors $n^a$ thereby  leading  to a condition on  the \textit{background
metric.} 
(Of course, one can specify any null vector $n^a(x)$ by giving its components $f^A(x)\equiv n^ae_a^A$ with respect to fixed set of basis vectors $e_a^A$ with $e_A^be_b^B=\delta^B_A$ etc so that $n^a=f^Ae_A^a$. So the class of all null vectors can be mapped to the scalar functions $f^A$ with the condition $f_Af^A=0$.)
Obviously, the functional in \eq{ent-func-2} must be rather special to accomplish this and one needs to impose  restrictions on  $P_{ab}^{\ph{a}\ph{b}cd}$ (and $T_{ab}$ though that condition turns out to be trivial) to achieve this.\index{action principle!thermodynamic approach}

It turns out --- as we shall see below --- that two conditions are sufficient
to ensure this.
First, the tensor $P_{abcd}$ should
have the same algebraic symmetries as the Riemann tensor $R_{abcd}$
of the $D$-dimensional spacetime. 
This condition can be ensured if we define $P_a^{\phantom{a}bcd}$ as
\begin{equation}
P_a^{\phantom{a}bcd} = \frac{\partial L}{\partial R^a_{\phantom{a}bcd}}
\label{condd1}
\end{equation} 
where $L = L(R^a_{\phantom{a}bcd}, g^{ik})$ is some scalar. 
Second, I will postulate the condition:
\begin{equation}
\D_{a}P^{abcd}=0.
\label{ent-func-1}
\end{equation}
as well as $\D_{a}T^{ab}=0$ which is anyway satisfied by any matter energy-momentum tensor.

\ha Can you give some motivation for these conditions ?

\me 
As regards \eq{condd1}, the motivation will become clearer later on. Basically, I will show that this approach leads to the same field equations as the one with $L$ as gravitational Lagrangian in the conventional approach (That is why I have used the symbol $L$ for this scalar!).

One possible motivation for \eq{ent-func-1} arises from fact that it will ensure  the field equations do not contain any derivative higher than second order of the metric. Another possible interpretation arises from the 
 analogy introduced earlier.
If you think of $n^a$ as analogous to deformation field in elasticity, then, in theory of elasticity \cite{landau7} one usually postulates the form of the thermodynamic potentials which are quadratic in first derivatives of $n_a$. The coefficients of this term will be the elastic constants. Here the coefficients are $P^{abcd}$ and you may want to think of \eq{ent-func-1} as saying the `elastic constants of spacetime solid' are actually `constants'. But nothing depends on this picture. In fact, I will show you later how this condition in \eq{ent-func-1} can be relaxed. 

\ha Interesting. You claim extremizing \eq{ent-func-2} in this context with respect to all $n^a$ leads to an equation constraining the background metric. If so, this is a peculiar variational principle.

\me Let me show you how this arises. 
Varying the normal vector field $ n^a$ after adding a
Lagrange multiplier function $\lambda(x)$ for imposing the   condition
$ n_a\delta  n^a=0$, we get 
\begin{equation}
-\delta S = 2\int_\Cal{V}{d^Dx\sqrt{-g}
  \left(4P_{ab}^{\ph{a}\ph{b}cd}\D_c n^a\left(\D_d\delta n^b\right)
  - T_{ab} n^a\delta n^b - \lambda(x) g_{ab} n^a\delta n^b\right)}
  \label{ent-func-3}
\end{equation}
where we have used the symmetries of $P_{ab}^{\ph{a}\ph{b}cd}$ and
$T_{ab}$.  An integration by parts and the
condition $\D_dP_{ab}^{\ph{a}\ph{b}cd}=0$, leads to 
\begin{eqnarray}
-\delta
S&=& 2\int_\Cal{V}{d^Dx\sqrt{-g}\left[-4P_{ab}^{\ph{a}\ph{b}cd}
  \left(\D_d\D_c n^a\right) - ( T_{ab}+ \lambda g_{ab}) n^a\right]\delta n^b}\nonumber\\
  &&+8\int_{\dV}{d^{D-1}x\sqrt{h}\left[k_d
  P_{ab}^{\ph{a}\ph{b}cd}\left(\D_c n^a\right)\right]\delta n^b}
\,,
\label{ent-func-4}
\end{eqnarray}
where $k^a$ is the $D$-vector field normal to the boundary \dV\ and
$h$ is the determinant of the intrinsic metric on \dV.  As usual, in order for
the variational principle to be well defined, we require that the
variation $\delta n^a$ of the  vector field should vanish on the
boundary. The second term in \eq{ent-func-4} therefore vanishes, and
the condition that $S[ n^a]$ be an extremum for arbitrary variations of
$ n^a$ then becomes  
\begin{equation}
2P_{ab}^{\ph{a}\ph{b}cd}\left(\D_c\D_d-\D_d\D_c\right) n^a
-( T_{ab}+\lambda g_{ab}) n^a = 0\,,
\label{ent-func-5}
\end{equation}
where we used the antisymmetry of $P_{ab}^{\ph{a}\ph{b}cd}$ in its
upper two indices to write the first term. The definition of the
Riemann tensor in terms of the commutator of covariant derivatives
reduces the above expression to
\begin{equation}
\left(2P_b^{\ph{b}ijk}R^a_{\ph{a}ijk} -  T{}^a_b+\lambda \delta^a_b\right) n_a=0\,, 
\label{ent-func-6}
\end{equation}
and we see that the equations of motion \emph{do not contain}
derivatives with respect to $n^a$ which is, of course, the crucial point. This peculiar feature arose because
of the symmetry requirements we imposed on the tensor
$P_{ab}^{\ph{a}\ph{b}cd}$. We  require that the condition in
\eq{ent-func-6} holds for \emph{arbitrary}  vector fields
$ n^a$. One can easily show\cite{aseementropy} that this requires
\begin{equation}
16\pi\left[ P_{b}^{\ph{b}ijk}R^{a}_{\ph{a}ijk}-\frac{1}{2}\delta^a_bL\right]=
 8\pi T{}_b^a +\Lambda\delta^a_b   
\label{ent-func-71}
\end{equation}
Comparison with \eq{genEab} shows that these are  precisely the field equations for  gravity (with a cosmological constant arising as an undetermined integration constant; more about this later) in a theory with Lagrangian $L$ when \eq{ent-func-2} is satisfied. That is,
 we have $2E_{ab}=T_{ab}+\lambda g_{ab}$ with
 \begin{equation}
E_{ab}=P_a^{\phantom{a} cde} R_{bcde}  - \frac{1}{2} L g_{ab};\quad
P^{abcd} \equiv \frac{\partial L}{\partial R_{abcd}}
\label{genEab1}
\end{equation}
The crucial difference between \eq{genEab} and  \eq{genEab1} 
is  that, the $E_{ab}$ in
\eq{genEab1} contains no derivatives of the metric higher than second order thereby leading to field equations which are second order in the metric. In contrast, \eq{genEab} can contain up to fourth order  derivatives of the metric.

\ha Let me get this straight. Suppose I start with a total Lagrangian $L(R_{abcd},g_{ab})+L_{matt}$, define a $P^{abcd}$ by \eq{condd1} ensuring it satisfies
\eq{ent-func-1}. Then I get certain field equations by varying the metric. You have just proved that I will get the \textit{same} field equations (but with a cosmological constant) if I start with the expression in \eq{ent-func-2}, maximize it with respect to $n^a$ and demand that it holds for all $n^a$. The maths is clear but I have several doubts. To begin with, why does the maths work out ?! 

\me I will let you into the secret by doing it differently. Note that, using the constraints on $P^{abcd}$ I can prove the identity
\begin{eqnarray}
\label{details1}
4P_{ab}^{\ph{a}\ph{b}cd} \D_cn^a\D_dn^b&=&
4\D_c[P_{ab}^{\ph{a}\ph{b}cd} n^a\D_dn^b]-4n^aP_{ab}^{\ph{a}\ph{b}cd} \D_c\D_dn^b\nonumber\\
&=&4\D_c[P_{ab}^{\ph{a}\ph{b}cd} n^a\D_dn^b]-2n^aP_{ab}^{\ph{a}\ph{b}cd} \D_{[c}\D_{d]}n^b\nonumber\\
&=&4\D_c[P_{ab}^{\ph{a}\ph{b}cd} n^a\D_dn^b]-2n^aP_{ab}^{\ph{a}\ph{b}cd} R^b_{\phantom{b}icd}n^i\nonumber\\
&=&4\D_c[P_{ab}^{\ph{a}\ph{b}cd} n^a\D_dn^b]+2n^aE_{ai}n^i
\end{eqnarray} 
where the first line uses \eq{ent-func-1}, the second line uses the antisymmetry of $P_{ab}^{\ph{a}\ph{b}cd}$ in c and d, the third line uses the standard identity for commutator of covariant derivatives and the last line is based on 
\eq{genEab} when $n_an^a=0$ and \eq{ent-func-1} hold. Using this in the expression for
 $S$ in \eq{ent-func-2} and integrating the four-divergence term, I can write
\begin{equation}
S[n^a]=-\int_{\partial\Cal{V}}{d^{D-1}x k_c\sqrt{h}}
(4P_{ab}^{\ph{a}\ph{b}cd} n^a\D_dn^b)
-\int_\Cal{V}{d^Dx\sqrt{-g}}\left[(2E_{ab}-T_{ab})n^an^b\right]
\label{thetrick}
\end{equation}
So, when I consider variations ignoring the surface term I am effectively varying $(2E_{ab}-T_{ab})n^an^b$ with respect to $n_a$ and demanding that it holds for all $n_a$. That should explain to you why it leads to $(2E_{ab}=T_{ab})$ except for a cosmological constant.

\ha Ha! I see it. Of course, there is an ambiguity of adding a term of the form $\lambda(x)g_{ab}$ in the integrand of the second term in \eq{thetrick} leading to the final equation
$(2E_{ab}=T_{ab}+\lambda(x)g_{ab})$ but the Bianchi identity $\nabla_aE^{ab}=0$ along with
$\nabla_aT^{ab}=0$ will make $\lambda(x)$ actually a constant. 

\me Yes. Remember that. We will discuss cosmological constant issue separately in the end.

\ha I see that it also connects up with your previous use of $2E_{ab}n^an^b$ as some kind of gravitational entropy density. Your expression for gravitational entropy is actually
\begin{eqnarray}
\label{thetrick1}
S_{grav}[n^a]&=&-\int_\Cal{V}{d^Dx\sqrt{-g}}4P_{ab}^{\ph{a}\ph{b}cd} \D_cn^a\D_dn^b\\
&=&-\int_{\partial\Cal{V}}{d^{D-1}x k_c\sqrt{h}}
(4P_{ab}^{\ph{a}\ph{b}cd} n^a\D_dn^b)
-\int_\Cal{V}{d^Dx\sqrt{-g}}(2E_{ab}n^an^b)\nonumber
\end{eqnarray}
Written in this form you have bulk contribution (proportional to our old friend $2E_{ab}n^an^b$) and a surface contribution. When equations of motion hold, the bulk also get a contribution from matter which cancels it out leaving the entropy of a region $\Cal{V}$ to reside in its boundary  $\partial\Cal{V}$.
\me Yes. I need to think more about this. 
\ha Also, arising out of your letting me into the trick, I realize that I can now find an $S$ for any theory, even if \eq{condd1} does not hold. You just have to reverse engineer it starting from $(2E_{ab}-T_{ab})n^an^b$ as the entropy density and using the expression in \eq{genEab} for $E_{ab}$, right ? So why do you insist on \eq{condd1} ?

\me You are right, of course. If you start with $(2E_{ab}-T_{ab})n^an^b$ as the entropy density  (see Eq. 14 of first paper in Ref.~\cite{tp09papers}) and work backwards you will get for $S_{\rm grav}$ the expression:
\begin{eqnarray}
S_{\rm grav} &=& - 4 \int_V d^Dx\, \sqrt{-g}\, \left( P^{abcd} \nabla_c n_a \, \nabla_d n_b + (\nabla_d P^{abcd})n_b \nabla_cn_a\right.\nonumber\\
&&\qquad \left. + (\nabla_c \nabla_dP^{abcd}) n_a n_b\right)
\label{generalS}
\end{eqnarray} 
Varying this with respect to $n^a$ will then lead to the correct equations of motion and --- incidentally --- the same surface term.

While one could indeed work with this more general expression, there are four reasons to prefer the imposition of the condition in  \eq{condd1}.
First,
 it is clear from \eq{genEab} that when $L$ depends on the curvature tensor and
the metric, $E_{ab}$ can depend up to the fourth derivative of the metric if \eq{condd1}
is not satisfied. But when we impose \eq{condd1} then we are led to field equations
which have, at most, second derivatives of the metric tensor -- which is  a desirable feature.
 Second, 
 as we shall see   below,  with that condition we can actually determine the form of $L$; it turns out that in $D=4$, it uniquely selects Einstein's theory, which  is probably a nice feature. In higher dimensions, it picks out a very geometrical extension of Einstein's theory in the form of \LL\ theories.
Third,
it is difficult to imagine why the terms in \eq{generalS} should occur with very specific coefficients. In fact, it is not clear  why we cannot have derivatives of $R_{abcd}$ in $L$, if the derivatives of $P_{abcd}$ can occur in the expression for entropy. 
Finally, 
if we take the idea of elastic constants being constants, then one is led to \eq{condd1}. 
None of these rigorously exclude the possibility in \eq{generalS} and in fact
this model has been
 explored  recently \cite{sfwu}.

\ha So far we have not fixed $P^{abcd}$ so we have not fixed the theory. How does \eq{condd1} allow you to do this?

\me
In a
complete theory, the explicit form of $P^{abcd}$ will be determined by the
long wavelength limit of the microscopic theory just as the elastic
constants can --- in principle --- be determined from the microscopic
theory of the lattice. In the absence of such a theory, we need to determine $P^{abcd}$ by general considerations. Essentially we need to determine scalar $L$ built from curvature tensor and the metric which satisfies the
the constraint $\nabla_a (\partial L/\partial R_{abcd})=0$. This problem can be completely solved the result is the Lagrangian of a \LL\ theory. Such an $L$ can be written  as a sum of terms, each involving products of  curvature tensors with the $m-$th term being a product of $m$ curvature tensors leading to
\begin{equation}
{L} = \sD{c_m\LDm}\,~;~{L}_{(m)} = \frac{1}{16\pi}
2^{-m} \Alt{a_1}{a_2}{a_{2m}}{b_1}{b_2}{b_{2m}}
\Riem{b_1}{b_2}{a_1}{a_2} \cdots \Riem{b_{2m-1}}{b_{2m}}{a_{2m-1}}{a_{2m}}
\,,  
\label{twotwo}
\end{equation}
where the $c_m$ are arbitrary constants and \LDm\ is the $m$-th
order \LL\ Lagrangian.
The $m=1$ term is proportional to $\delta^{ab}_{cd}R^{cd}_{ab} \propto R$ and leads
to Einstein's theory. 
It is conventional to take $c_1 =1$ so that
the ${\ensuremath{{L}_{(1)}}}$,  reduces to $R/16\pi$.
The normalizations for $m>1$ are somewhat arbitrary for individual \LDm\ since the $c_m$s 
are unspecified at this stage.
The $m=2$ term gives rise to what is known as Gauss-Bonnet theory.
Because of the determinant tensor, it is obvious that in any given dimension $D$ we can only have $K$ terms where 
$2K\leq D$.
It follows that, if $D=4$, then only the $m=1, 2$ are non-zero.
Of these, the Gauss-Bonnet term (corresponding to $m=2$) gives, on variation of the
action, a vanishing bulk contribution in $D=4$.
(In dimensions $D=5 $ to  8, one can have both the Einstein-Hilbert term
and the Gauss-Bonnet term and so on.)
Equivalently, 
the $P^{abcd}$ can be expressed as a series in the
 powers of  derivatives of
the metric as:
\begin{equation}
P^{abcd} (g_{ij},R_{ijkl}) = c_1\,\overset{(1)}{P}{}^{abcd} (g_{ij}) +
c_2\, \overset{(2)}{P}{}^{abcd} (g_{ij},R_{ijkl})  
+ \cdots \,,
\label{derexp}
\end{equation} 
where $c_1, c_2, \cdots$ are coupling constants.   The lowest order
term depends only on the metric with no derivatives. The next
term depends (in addition to metric) linearly on curvature tensor and the next one will be quadratic in curvature etc. 

Let us  take a closer look at the structure which is emerging. The lowest order term in \eq{derexp} (which  leads to Einstein's theory) is
\begin{equation}
\overset{(1)}{P}{}^{ab}_{cd}=\frac{1}{16\pi}
\frac{1}{2} \delta^{ab}_{cd} =\frac{1}{32\pi}
(\delta^a_c \delta^b_d-\delta^a_d \delta^b_c)
  \,.
\label{pforeh}
\end{equation}
To the lowest order, when we use \eq{pforeh} for $P_{b}^{\ph{b}ijk}$, 
  the \eq{ent-func-71} 
 reduces to Einstein's equations.
 The corresponding gravitational entropy functional is:
 \begin{equation}
S_{GR}[n^a]=\int_\Cal{V}\frac{d^Dx}{8\pi}
   \left(\D_an^b\D_bn^a - (\D_cn^c)^2 \right)
\end{equation}
Interestingly, the integrand in $S_{GR}$ has the $Tr(K^2)-(Tr K)^2$ structure. If we think of the $D=4$ spacetime being embedded in a sufficiently large k-dimensional \textit{flat} spacetime we can obtain the same structure using the Gauss-Codazzi equations relating the (zero) curvature of k-dimensional space with the curvature of spacetime.
As mentioned earlier, one can express any vector field $n^a$ in terms of a set of basis vector fields $n^a_A$. Therefore, one can equivalently think of the functional $S_{\rm GR}$ as given by 
 \begin{equation}
S_{\rm GR}[n^a_A]=\int_\Cal{V}\frac{d^Dx}{8\pi}
   \left(\D_an^b_I\D_bn^a_J - \D_cn^c_I\D_an^a_J \right)P^{IJ}
\end{equation}
where $P^{IJ}$ is a suitable projection operator. It is not clear whether the embedding approach leads to any better understanding of the formalism; in particular, it does not seem to generalize in a natural fashion to \LL\ models.

 The next order term (which arises from  the  Gauss-Bonnet Lagrangian) is:
\begin{equation}
\overset{(2)}{P}{}^{ab}_{cd}= \frac{1}{16\pi}
\frac{1}{2} \delta^{ab\,a_3a_4}_{cd\,b_3\,b_4}
R^{b_3b_4}_{a_3a_4} =\frac{1}{8\pi} \left(R^{ab}_{cd} -
         G^a_c\delta^b_d+ G^b_c \delta^a_d +  R^a_d \delta^b_c -
         R^b_d \delta^a_c\right) 
\label{pingone}
\end{equation} 
and similarly for all the  higher orders terms. None of them can contribute in $D=4$ so we get Einstein's theory as the unique choice if we assume $D=4$. If we assume that $P^{abcd}$ is to be built \textit{only} from the metric, then this choice is unique in all $D$. 

\ha You originally gave a motivational argument as to why this $S$ should be thought of as entropy. As far as the variational principle is concerned, this identification does not seem to play a crucial role.

\me It does rather indirectly. To see this, you only need to consider the form of $S$ when the equations of motion are satisfies. First of all, \eq{thetrick} shows that when the equations of motion holds the total entropy of a bulk region is entirely on its boundary, which is nice. Further
if you evaluate this boundary term
\begin{equation}
-S|_{\rm on-shell}=4\int_{\dV}{d^{D-1}xk_a\sqrt{h}\,\left(P^{abcd}n_c\D_bn_d\right)}
\label{on-shell-2}
\end{equation} 
(where we have manipulated a few indices using the symmetries of
$P^{abcd}$) 
in the case of a \textit{stationary} horizon which can be locally approximated as Rindler spacetime, one gets exactly the Wald entropy of the horizon \cite{aseementropy}. This is one clear reason as to why we can think of $S$ as entropy.

 \ha Is this entropy positive definite? Do you worry about that?
 
 \me I don't worry about that (yet!). In $D=4$, I can prove that the on-shell entropy is positive definite. But if one is dealing with a \LL\ model with horizons attributed with Wald entropy, it is known that \cite{entneg} even on-shell entropy 
will not be positive definite for all range of parameters. May be this will put additional restrictions on the kind of gravitational theories which are physically reasonable.
(This approach has uncovered several other issues related to entropy, quasi-normal modes etc. and even a possibility of entropy being quantized \cite{entropyideas} but all that will take us far afield.)
At present these questions are open.

 \ha Usually in an action principle one varies all the degrees of freedom in any order one chooses. But in your extremum principle, we are expected to vary only $n^a$. How would I get equations of motion for matter in this approach?
 
 \me That is not a problem. 
 At the classical level, the field equations are already contained in the condition $\nabla_a T^{ab} =0$ which I impose (with an intriguing interpretation, which you may not want to buy, that this is the constancy of elastic constants!) If you want to do quantum field theory in a curved spacetime, you can again use these field equations in the Heisenberg picture. The only question is when you insist that you need to do a path integral quantizations of the matter fields. Then, 
 you have to, of course, vary $n^a$ first  and get the classical equations for gravity
 because the expression in \eq{ent-func-2} is designed as an entropy functional.  But after you have done that and written down
 the field equations for gravity, you can do the usual variation of matter Lagrangian in a given curved spacetime and get the standard equations \cite{aseementropy}.

\section{Comparison with the conventional perspective and further comments}\label{sec:comaparison}

 \ha I also notice that while the vector field $n_a$ in LIF, to the lowest order, has no bulk dynamics,  if you consider the integral over the Lagrangian in a small region in LIF, you will get a surface contribution. That seems strange, too.
  
  \me Not really. I am not surprised by $S$ picking up just a surface contribution because --- even in the conventional approach --- Einstein-Hilbert action   is holographic in a specific sense of the word \cite{cc1} and does exactly that. 

\ha May be this is good time to sort this out. You mentioned earlier that the democracy of observers and their right to do physics in spite of the existence of horizons has something to do with the holography of action. I have no idea what you are talking about here!

 \me Let me elaborate. We said that there should exist a mechanism which will encode
  the information in the region $\mathcal{V}$ which is inaccessible to a particular observer at the boundary $\partial \mathcal{V}$ of that region\cite{tpapoorva}. 
 One possible way of ensuring this is to add a suitable boundary term to the action principle which will provide additional information content for observers who perceive a horizon. Such a procedure leads to three immediate consequences.
  
  First, if the theory is generally covariant, so that observers with horizons (like, for example, uniformly accelerated observers using a Rindler metric) need to be 
 accommodated in the theory, such a theory \textit{must} have an action functional that contains
  a surface term. The generally covariant action in 
   Einstein's theory did contain a surface term. The present approach explains
   the logical necessity for such a surface term in a generally covariant theory which was not evident in the 
   standard approach.
   
   \ha That's interesting. You are now claiming that there is a connection between the following three facts. (1) The theory for gravity is built from a generally covariant Lagrangian. (2) In a geometrical theory of gravity, horizons are inevitable but general covariance demands that all observers have an equal right to describe physics. (3) Observers whose information is blocked by a horizon should still be able to 
   somehow get around this fact with the information encoded on the boundary. Therefore, the Lagrangian must have a boundary term. Viewed this way, it appears natural that the only generally covariant scalar Lagrangian proportional to  $R$ leads to a surface term in the action. But how does the surface term know what it is going on in the bulk? 
   
   \me That is the second point. If the surface term has to  encode the information which is blocked by the horizon,
   then there must exist a simple relation between the bulk term
   and surface term in the action and hence you cannot choose just any scalar.  This is indeed the case for the Einstein-Hilbert action;
    there is a 
peculiar (unexplained) relationship between $L_{\rm bulk}$ and $L_{\rm sur}$:
\begin{equation}
    \sqrt{-g}L_{sur}=-\partial_a\left(g_{ij}
\frac{\partial \sqrt{-g}L_{bulk}}{\partial(\partial_ag_{ij})}\right)
\label{pecurrel}
\end{equation}
This shows that the Einstein-Hilbert gravitational action is `holographic' with the same information being coded in both the bulk and surface terms.  

In fact, in any local region around an event, it is the surface term which contributes to the action at the lowest order.
In the neighborhood of any event, the Riemann normal coordinates in which $g\simeq \eta + R\ x^2, \Gamma\simeq R\ x$. In the gravitational Lagrangian $\sqrt{-g}R\equiv\sqrt{-g}L_{bulk}+\partial_a P^a$  with $L_{bulk}\simeq \Gamma^2$ and $\partial P\simeq \partial \Gamma$, the $L_{bulk}$ term vanishes in this neighborhood while
$\partial_aP_b\simeq R_{ab}$ leading to
\begin{equation}
\int_{\mathcal{V}} d^4x R\g \approx \int_{\mathcal{V}} d^4x \partial_a P^a
 \approx \int_{\partial\mathcal{V}} d^3x n_a P^a
\end{equation} 
showing that in a small region around the event in the Riemann normal coordinates, gravitational action can be reduced to a pure surface term.

\ha So in this perspective, you also expect the surface term to be related to the information content blocked by the horizon, right?

   \me Indeed. That is the third point.
   If the surface term encodes information which is blocked by the horizon, then
   it should actually lead to the entropy of the horizon. In other words, we should be able
   to compute the horizon entropy by evaluating the surface term. This is indeed true
   and can be easily demonstrated \cite{cc1}. 
     The surface term does give the horizon entropy 
    for any metric for which near-horizon geometry has the Rindler form.

This explains another deep mystery in the conventional approach. 
 In the usual  approach, we \textit{ignore} the surface term completely
(or cancel it with a counter-term) and obtain the field equation
from the bulk term in the action. Any solution to the field equation obtained 
by this procedure is logically independent of the nature of the surface term.
But we find that when the \textit{surface term} (which was ignored) is evaluated at the horizon that arises
in any given solution, it does correctly give the entropy of the horizon!
This is possible only because there is a  relationship, given by \eq{pecurrel},  
between the 
surface term and the bulk term which is again an unexplained feature in the conventional
approach to gravitational dynamics.
Since the surface term has the thermodynamic interpretation as the entropy of horizons,
 and is related holographically to the bulk term, we are again led to 
  an indirect connection between spacetime dynamics and horizon thermodynamics.
  
\ha I agree these results are extremely mysterious in the conventional approach, now that you brought it up. I have not seen \eq{pecurrel} mentioned, let alone discussed in any work (other than yours, of course.) I presume this is one of what you call `algebraic accidents'.
But if your ideas about \LL\ theory being the natural candidate in D dimensions then the same `algebraic accident' should occur in \LL\ theories as well, right?

\me Yes. In fact it does --- which is gratifying --- and acts as a nontrivial consistency check on my alternative perspective. One can show that the surface and bulk terms of all \LL\ theories satisfy an equation similar to  \eq{pecurrel}. Of course, since it wasn't noticed for Hilbert action, nobody bothered about \LL\ action till we \cite{ayan} unearthed it. 

One can provide a simple, yet very general, proof of the connection between entropy and surface term in action in any static spacetime. Such a spacetime will have a Killing vector $\xi^a$ and a corresponding Noether current. Taking the $J^0$ component of \eq{current1} and writing $J^0 = \nabla_b J^{0b}$
we obtain 
\begin{equation}
L = \frac{1}{\sqrt{-g}} \partial_\alpha \left( \sqrt{-g}\, J^{0\alpha}\right) - 2 E^0_0
\label{strucL}
\end{equation} 
Only spatial derivatives contribute in the first term on the right hand side when the 
spacetime is static. This relation shows that the action obtained by integrating $L\sqrt{-g}$ will generically have a surface term related to $J^{ab}$ (In Einstein gravity \eq{strucL} will read as $L=2R^0_0-2G^0_0$; our result generalises the fact that $R^0_0$ can be expressed as a total divergence in static spacetimes.) This again illustrates, in a very general manner, why the surface terms in the action functional
lead to horizon entropy. In fact \eq{strucL} can be integrated to show that in any static spacetime with a bifurcation horizon, the action can be interpreted as the free energy which generalises a result known in Einstein gravity to \LL\ models.

\ha What are the other key algebraic accidents in the conventional approach which your perspective throws light on ?

\me There are several but let me describe one which is really striking (and was first discussed in ref. \cite{tdsingr}).
Consider a static, spherically symmetric horizon, in a spacetime described by a metric:
\begin{equation}
ds^2 = -f(r) c^2 dt^2 + f^{-1}(r) dr^2 + r^2 d\Omega^2. \label{spmetric}
\end{equation}
Let the location of  the horizon be given by the simple zero of the function $f(r)$, say at $r=a$. The Taylor series expansion of $f(r)$ near the horizon $f(r)\approx f'(a)(r-a)$ shows that the metric reduces to the Rindler metric near the horizon in the $r-t$ plane  with the surface gravity $\kappa = (c^2/2) f'(a)$. Then, an analytic continuation to imaginary time allows us to identify the temperature associated with the horizon to be 
\begin{equation}
k_BT=\frac{\hbar c f'(a)}{4\pi}
\label{hortemp1}
\end{equation} 
where we have introduced the normal units. 
The association of temperature in \eq{hortemp1} with the metric in \eq{spmetric} only requires the conditions $f(a)=0$ and $f'(a)\ne 0$.
The discussion so far did not assume anything about the dynamics of gravity or Einstein's field equations.  

We shall now take the next step and write down the
Einstein equation for the metric in \eq{spmetric}, which is given by 
$(1-f)-rf'(r)=-(8\pi G/c^4) Pr^2$ where $P = T^{r}_{r}$ is the radial pressure. When evaluated on the horizon $r=a$ we get the result:
\begin{equation}
\frac{c^4}{G}\left[\frac{1}{ 2} f'(a)a - \frac{1}{2}\right] = 4\pi P a^2
\label{reqa}
\end{equation}
If we now consider two solutions to the Einstein's equations differing infinitesimally in the parameters such that horizons occur at two different radii $a$ and $a+da$,
then multiplying the \eq{reqa} by $da$, we get: 
\begin{equation}
\frac{c^4}{2G}f'(a) a da - \frac{c^4}{2G}da = P(4\pi a^2 da)
\label{reqa1}
\end{equation}
The right hand side is just $PdV$ where $V=(4\pi/3)a^3$ is what is called the areal volume which is the relevant quantity when we consider the action of pressure on a surface area. In the first term, we note that $f'(a)$ is proportional to horizon temperature in \eq{hortemp1}. Rearranging this term slightly
and introducing a $\hbar$ factor \textit{by hand} into an otherwise classical equation
to bring in the horizon temperature, we can rewrite \eq{reqa1} as
\begin{equation}
   \underbrace{\frac{{{\hbar}} cf'(a)}{4\pi}}_{\displaystyle{k_BT}}
    \ \underbrace{\frac{c^3}{G{{\hbar}}}d\left( \frac{1}{ 4} 4\pi a^2 \right)}_{
    \displaystyle{dS}}
  \ \underbrace{-\ \frac{1}{2}\frac{c^4 da}{G}}_{
    \displaystyle{-dE}}
 = \underbrace{P d \left( \frac{4\pi}{ 3}  a^3 \right)  }_{
    \displaystyle{P\, dV}}
\label{EHthermo}
\end{equation}
The labels below the equation indicate a natural --- and unique --- interpretation for each of the terms and the whole equation now becomes $TdS=dE+PdV$ allowing us to read off the expressions for entropy and energy:
\begin{equation}
 S=\frac{1}{ 4L_P^2} (4\pi a^2) = \frac{1}{ 4} \frac{A_H}{ L_P^2}; \quad E=\frac{c^4}{ 2G} a
    =\frac{c^4}{G}\left( \frac{A_H}{ 16 \pi}\right)^{1/2}
\end{equation}
where $A_H$ is the horizon area and $L_P^2=G\hbar/c^3$. The result shows that Einstein's equations can be re-interpreted as a thermodynamic
identity for a virtual displacement of the horizon by an amount $da$. 

\ha  
I suppose the uniqueness of the factor $P(4\pi a^2) da$, where $4\pi a^2$ is the 
proper area of a surface of radius $a$ in spherically symmetric spacetimes, 
implies that we cannot carry out the same exercise by multiplying \eq{reqa} by   some other arbitrary factor 
$F(a) da$ instead of just $da$ in a natural fashion. This, in turn, uniquely fixes both $dE$ and the 
combination $TdS$. The product $TdS$ is  classical and is independent of $\hbar$ and hence we can  determine $T$ and $S$ only within
a multiplicative factor. The only place you introduced $\hbar$ by hand is in
 using the Euclidean extension of the metric to fix the form of $T$ and thus $S$.  Right ?

\me Yes. With that I can remove the ambiguity
in the overall multiplicative factor. So, given the structure of the metric
in \eq{spmetric} and Einstein's equations, we can determine $T,S$ and $E$
uniquely.  The fact that $T\propto \hbar$ and $S\propto 1/\hbar$ is  analogous to the situation in classical thermodynamics  in contrast with statistical mechanics. The $TdS$ in thermodynamics is independent of Boltzmann's constant while statistical mechanics will lead to  $S\propto k_B$ and $T\propto1/k_B$.

\ha That is a bit mind-boggling. Usually, the rigorous way of obtaining the temperature of a horizon --- say, a black hole horizon --- is by studying a quantum field in the externally specified metric. \textit{You never need to specify whether the metric is a solution to Einstein's equations.} Now you are telling me that the same result arises \textit{without any reference
to an externally specified quantum field theory but on using Einstein's equations on the horizons}. How does the Einstein equations know that there is a temperature, entropy etc ?

\me Yes. That is the algebraic coincidence. More sharply stated, we have no explanation as to why an equation like \eq{EHthermo} should hold in classical gravity, if we take the conventional route. This strongly suggests that the association of entropy and temperature with a horizon
is quite fundamental and is actually connected with the dynamics (encoded in Einstein's equations)
of the gravitational field. The fact that quantum field theory in a spacetime
with horizon exhibits thermal behaviour should then be thought of as a \textit{consequence} 
of a more fundamental principle. 

\ha If so the idea should also have a more general validity. Does it ?

\me Yes. One can again show that the field equations of more general theories of gravity
(like in \LL\ models) also reduce to the same thermodynamic identity $TdS = dE +P dV$
when evaluated on the horizon. This has now been demonstrated \cite{tdsforlletc} for an impressively wide class of models
 like (i) the  stationary
axisymmetric horizons and (ii) evolving spherically symmetric horizons
in Einstein gravity, (iii) static spherically symmetric
horizons and (iv) dynamical apparent horizons in
Lovelock gravity, and (v) three dimensional BTZ black hole
horizons, (vi) FRW cosmological
models in various gravity
theories and (vii) even \cite{hwg} in the case Horava-Lifshitz Gravity. It is not possible to understand, in the conventional approach, why the field equations should encode information about horizon thermodynamics.

\ha This is a fairly strong argument in favour of a thermodynamic underpinning for the 
dynamics of gravity. But before I accept that in toto, I need to convince myself
that there is no simpler explanation for this result. I accept that none is given in the literature but how about the standard first law of black hole thermodynamics? Your relation looks similar to it so I wonder whether there is  a connection.

\me No. We are talking about very different things. 
In general,  
 in spite of the superficial similarity, \eq{EHthermo} is \textit{different}
from the conventional first law of black hole thermodynamics due to the presence of
$PdV$ term.  The difference  is easily seen, for example,
in the case of Reissner-Nordstrom black hole for which $T^r_r = P$ is non-zero due to the presence of nonzero electromagnetic energy-momentum tensor in the right hand 
side of Einstein's equations. If a
\textit{chargeless} particle of mass $dM$ is dropped into a
Reissner-Nordstrom black hole, then 
the standard first law of black hole thermodynamics 
 will give $TdS=dM$. But in \eq{EHthermo}, the energy term,
defined as  $E\equiv a/2$, changes by $dE= (da/2)
=(1/2)[a/(a-M)]dM\neq dM$. 
 It is easy to see, however, that for the Reissner-Nordstrom black hole, the combination 
 $dE+PdV$  is precisely
equal to $dM$ making sure $TdS=dM$. So we need the $PdV$ term to get
$TdS=dM$ from \eq{EHthermo}  when a \textit{chargeless} particle is dropped into a
Reissner-Nordstrom black hole. More generally, if $da$ arises due to
changes $dM$ and $dQ$, it is easy to show  that \eq{EHthermo} gives
$TdS=dM -(Q/a)dQ$ where the second term arises from the electrostatic
contribution. 
This ensures that \eq{EHthermo} is perfectly consistent
with the standard first law of black hole dynamics in those contexts
in which both are applicable but  $dE \ne dM$ in general.
You would have also realized that the way 
\eq{EHthermo}
was derived is completely local and quite different from the way one obtains
first law of black hole thermodynamics.

\ha Yes, I see that. It appears that gravitational field equations and their
solutions with horizons has a deeper connection with thermodynamics than is
apparent. In fact, I believe you would claim the thermodynamic perspective
is more fundamental than the field equations describing gravity.

\me Precisely. That is why I spent lot of time explaining the thermodynamic
motivation  in Sections \ref{sec:lro} and \ref{sec:reinterpret} while I could have derived the field equations just by
extremizing the expression in   \eq{ent-func-2}. But in a way everything else is 
just motivational if you are willing to accept the perspective based on 
\eq{ent-func-2} as fundamental.

\ha That brings up the question you promised a discussion on. What about the
cosmological constant \cite{ccreview}? In the conventional approach, one introduces it as a 
term in the gravitational Lagrangian. You don't have any such term in \eq{ent-func-2}
but nevertheless the cosmological constant appears in your final equations!

\me Yes and I would claim that this is another very attractive feature of this new perspective.
 In the standard approach,
one starts with an action 
\begin{equation}
\mathcal{A}_{\rm tot}= \int d^Dx\, \g\, (  L_{\rm grav} + L_m)
\end{equation} 
and varies (i) the matter degrees of freedom to obtain the equations of motion for
matter and (ii) the metric $g^{ab}$ to obtain the field equations of gravity.
The equations of motion for matter remain invariant if one adds a constant,
say, $-\rho_0$ to the matter Lagrangian, which is equivalent to adding a constant
$\rho_0$ to the Hamiltonian density of the matter sector. 
Physically, this symmetry reflects the fact that the zero level 
of the energy is arbitrary in the matter sector and can be set to any value
without leading to observable consequences.
However, gravity breaks
this symmetry which the matter sector has.
A shift $L_m \to L_m - \rho_0$ will change the energy-momentum tensor $T^a_b$ which
acts as the source of gravity by a term proportional to  $\rho_0 \delta^a_b$. Therefore, having a nonzero baseline for energy density of matter
is equivalent to a theory with cosmological constant which --- in turn --- will lead to
observable consequences. If we interpret the evidence for dark energy in the 
universe (see ref. \cite{sn}; for a critical look at data, see ref. \cite{tptirthsn1} and references therein)
as due to the cosmological constant,\index{cosmological constant} then its value has to be
fine-tuned to enormous \footnote{This is, of course, the party line. But it might help to get some perspective on how enormous, the `enormous' really is. To begin with note that, the sensible particle physics
  convention considers ratios of length (or energy) scales and not their \textit{squares} as cosmologists are fond of doing. This leads to
  $(L_P/L_\Lambda) \sim 10^{-61}$ instead of the usual $\Lambda L_P^2\sim 10^{-122}$. In standard model of particle physics the ratio between Planck scale to neutrino mass scale is $10^{19}GeV/10^{-2}eV\sim 10^{30}$ for which we have no theoretical explanation. So when we worry about the fine tuning of cosmological constant without expressing similar worries about standard model of particle physics, we are essentially assuming that $10^{30}$ is not a matter for concern but $10^{61}$ is. This subjective view is defensible but needs to be clearly understood.} accuracy to satisfy the observational constraints.
It is not clear why a particular parameter in the low energy matter sector has to be fine-tuned in such a manner. 

In the alternative perspective described here,
the functional in \eq{ent-func-2} is clearly invariant under the shift $L_m \to L_m - \rho_0$ or equivalently, $T_{ab} \to T_{ab} + \rho_0 g_{ab}$, 
since it only introduces a term $-\rho_0 n_a n^a =0$ for any null vector $n_a$.
In other words, one \textit{cannot} introduce the cosmological constant
as  a low energy parameter in the action in this approach. We saw, however, 
that the cosmological constant reappears an \textit{an integration constant} when the equations 
are solved. The integration constants which appear in a particular solution  have a completely different conceptual status compared to the parameters which appear in the action describing
the theory. It is much less troublesome to choose a fine-tuned value for a particular integration constant in the theory if observations require us to do so.
From this point of view, the cosmological constant problem is considerably less severe
when we view gravity from the alternative perspective.

\ha I suppose you succeed in having the extra symmetry under the shift 
$T_{ab} \to T_{ab} + \rho_0 g_{ab}$ because you are not treating metric  as a 
dynamical variable, right?

\me Right. In fact one can state a stronger result \cite{gr06}.
Consider any model of gravity satisfying the following three conditions: (1)
The metric  is varied in a local action to obtain the equations of motion. (2) We demand full general covariance of the equations of motion. (3) The equations of motion for matter sector is invariant under the addition of a constant to the matter Lagrangian.  Then, we can prove `no-go' theorem that the \cc\ problem cannot be solved in such model \cite{gr06}. The proof is elementary. Our demand (2) of general covariance requires the matter action to be an integral over $\mathcal{L}_{matter}\sqrt{-g}$. 
The demand (3) now allows us to add a constant $\Lambda$, say, to $\mathcal{L}_{matter}$ leading to a coupling $\Lambda\sqrt{-g}$ between $\Lambda$ and the metric $g_{ab}$. By our demand (1), when we vary $g_{ab}$ the theory will couple to $\Lambda$ through a term proportional to $\Lambda g_{ab}$ thereby introducing an arbitrary \cc\ into the theory.

The power of the above `no-go theorem' lies in its simplicity! It clearly shows that we cannot solve \cc\ problem unless we drop one of the three demands listed in the above paragraph. Of these, we do not want to sacrifice general covariance encoded in (2); neither do we have a handle on low energy matter Lagrangian so we cannot avoid (3). So the only hope we have is to introduce an approach in which gravitational field equations are obtained from varying some degrees of freedom other than $g_{ab}$ in a maximization principle.
This suggests that the so called cosmological constant problem has its roots in our misunderstanding of the nature of gravity.

\ha I thought that
any spin-2 long range field 
$h_{ab}$ (arising, for example, in the 
linear perturbation around flat spacetime through $g_{ab}= \eta_{ab} + h_{ab}$)
obeying 
 principle of equivalence has to generically couple to 
$T_{ab}$ through a term in the action $T^{ab} h_{ab}$. But in your model, this does not
seem to happen. 

\me That's correct. 
It is sometimes claimed that a spin-2 graviton in the linear limit \textit{has to} couple to $T_{ab}$ in a universal manner, in which case, one will have the graviton coupling to the cosmological constant. In our approach, the linearized  field equations for the spin-2 graviton field $h_{ab}=g_{ab}-\eta_{ab}$, in a suitable gauge, will be $(\square h_{ab} -T_{ab})n^an^b=0$ for all null vectors $n^a$. This equation is still invariant under $T_{ab} \to T_{ab} + \rho_0 g_{ab}$ showing that the graviton does \textit{not} couple to cosmological constant.

\ha But can you predict the observed  value of the \cc\ in your approach?

\me Alas, no. But I claim providing a mechanism in which the \textit{bulk cosmological constant
decouples from gravity} is a major step forward. If the \cc\ was strictly zero, my perspective has a natural explanation for it --- which no one else had! It was always thought
that this should arise from some unknown symmetry and I have provided you with a model which has such symmetry. I believe the small value of the observed \cc\ arises from non-perturbative quantum gravitational effects at the next order, but I don't have a fully satisfactory model. (See, however, Ref. \cite{ccsol}.)

\section{Summary and outlook}

\ha We have covered a lot of ground some of which is purely technical while the rest are
conceptual or interpretational. In your mind the distinction may be unimportant
but others will react differently to results which can be rigorously proved compared to
interpretational aspects, however elegant the latter may be. May be you would care to separate
them out and provide a summary?

\me Fine. 
From a purely algebraic point of view, without bringing in any physical interpretation or motivation, we can prove the following mathematical results: 
\begin{itemize}
\item
Consider a functional of null vector fields $n^a(x)$ in an arbitrary spacetime given by
\eq{ent-func-2} [or, more generally, by \eq{generalS}]. Demanding that this functional is an extremum for all null vectors $n^a$ leads to the field equations for the background geometry given by $(2E_{ab}-T_{ab})n^an^b=0$ where $E_{ab}$ is given by \eq{genEab1} [or, more generally, by \eq{genEab}]. Thus field equations in a wide class of theories of gravity can be obtained from an extremum principle without varying the metric as a dynamical variable.
\item
These field equations are invariant under the transformation 
$T_{ab} \to T_{ab} + \rho_0 g_{ab}$, which 
relates to the freedom of introducing a  \cc\ as an integration constant in the theory. Further, this symmetry forbids the inclusion of a cosmological constant  term in the variational principle by hand as a low energy parameter. \textit{That is, we have found a symmetry which makes the bulk \cc\ decouple from the gravity.} When linearized around flat spacetime, the graviton inherits this symmetry and does not couple to the \cc .
\item
On-shell, the functional in \eq{ent-func-2} [or, more generally, by \eq{generalS}] contributes only on the boundary of the region. When the boundary is a horizon, this terms gives precisely the Wald entropy of the theory.
\end{itemize}

\ha It is remarkable that you can derive not only Einstein's theory uniquely in $D=4$ but even \LL\ theory in $D>4$ from an extremum principle involving the null normals \textit{without varying $g_{ab}$ in an action functional!}. I also see from \eq{details1}
that in the case of Einstein's theory, you have an Lagrangian $n^a(\nabla_{[a}\nabla_{b]})n^b$ for a vector field $n^a$ which becomes vacuous in flat spacetime in which covariant derivatives become partial derivatives. There is clearly no dynamics in $n^a$ but they do play a crucial role. 
So I see that you can get away without ever telling me what the null vectors $n^a$ actually means because they disappear from the scene after serving their purpose. While this may be mathematically clever, it is very unsatisfactory physically. You may not have a rigorous model for these degrees of freedom but what is your picture?

\me My picture is made of the following ingredients, each  of which seems reasonable but far from having rigorous mathematical justification at this stage.

\begin{itemize}
\item
Assume that the spacetime is endowed with certain microscopic degrees of freedom capable of 
exhibiting thermal phenomena. This is just the Boltzmann paradigm: \textit{If one can heat it, it
must have microstructure!}; and one can heat up a spacetime. 
\item
Whenever a class of observers perceive a horizon, they are ``heating up the spacetime''
and the  degrees of freedom close to a horizon  participate  
in a very \textit{observer dependent} thermodynamics. 
Matter which flows close to the horizon
(say, within a few Planck lengths of the horizon) transfers energy to these microscopic, near-horizon, degrees of freedom \textit{as far as the observer who sees the horizon is concerned}. Just as  entropy of a normal system at temperature $T$ 
will change by $\delta E/T$ when we transfer to it an energy $\delta E$, here also an entropy change will occur. (A freely falling observer in the same neighbourhood, of course, will deny all these!)
\item
We proved that when the field equations of gravity hold, one can interpret this entropy change in  a purely geometrical manner involving the Noether current.
From this point of view, the  normals $n^a$ to local patches of null surfaces are related to the (unknown) degrees of freedom that can participate in the thermal phenomena involving the horizon. 
\item
Just as demanding the validity of special relativistic laws with respect to all freely falling observers leads to the kinematics of gravity, demanding the local entropy balance in terms of the thermodynamic variables as perceived by local Rindler observers leads to the field equations of gravity in the form $(2E_{ab}-T_{ab})n^an^b=0$. 
 
\end{itemize}

\ha It is an interesting picture but is \textit{totally observer dependent}, right? A local Rindler
observer or an observer outside a black hole horizon might attribute all kinds of thermodynamics and entropy changes to the horizons she perceives. But an inertial observer
or an observer falling through the Schwarzschild horizon will see none of these phenomena.

\me Exactly. I claim we need to accept the fact that a whole lot of thermodynamic phenomena needs to be now thought of as observer dependent. For example, if you throw some hot matter on to a Schwarzschild black hole, then when it gets to a few Planck lengths away from the horizon and hovers around it, I expect it to interact with the microscopic horizon degrees of freedom \textit{as far as an outside observer is concerned}.
After all, such an observer would claim that all matter stays arbitrarily close but outside the horizon for all eternity. A freely falling observer through the horizon will have a completely different picture but we have learnt to live with this dichotomy as far as elementary kinematics goes. I think we need to do the same as regards thermodynamics and quantum processes. 

\ha In a way, every key progress in physics involved realizing that something we thought as absolute is  not absolute. With special relativity it was the flow of time and with general relativity it was the concept of global inertial frames and when we brought in quantum fields in curved spacetime it was the notion of particles. 

\me And the notion of temperature, don't forget that. We now know that the temperature attributed to even vacuum state depends on the observer. We need to go further and integrate the entire thermodynamic machinery --- involving highly excited semi-classical states, say, cups of tea with (what we believe to be) ``real'' temperature --- 
 with this notion of LRFs having their own temperature. I don't think this has been done in a satisfactory manner yet \cite{tpinprogress}.

\ha So, what next? What is the ``to do list''?

\me  To name a few which comes to ones mind,  I can list them as  technical ones  and conceptual ones. On the technical side: 

(i) It would be nice to make the notion of LRF and the horizon a bit more rigorous. For example, the idea of an approximate Killing vector could be made more precise and one might like to establish the connection between locality, that is apparent in the Euclidean sector and the causality, which is apparent in the Lorentzian section that has the light cones. By and large, one would like to make rigorous the use of 
LRFs  by, say, computing the next order corrections. 

(ii) A lot more can be done to clarify the observer dependence of the entropy. (The study of horizon thermodynamics makes one realize that one does not really quite understand what entropy is!). It essentially involves \textit{exact} computation of  $(S_1 - S_0)$
where $S_1 $ and $S_0$ are the entropies attributed to the excited and ground state by a Rindler observer. This should throw more light on the expression for $\delta S$ used in \eq{delS}. In particular, it would be nice to have a detailed model which
shows why $\delta S$ involves the combination of $\delta E$ of matter and
$T$ of the horizon. One would then use these insights to understand why \eq{Smatt}
actually represents the relevant entropy functional for matter for arbitrary $T_{ab}$. 

(iii) It will be nice to have a handle on the positivity or otherwise of the entropy
functional used in \eq{ent-func-2}. 

These technical issues, I believe can be tackled in more or less straightforward manner, 
though the mathematics can be fairly involved. But as I said, they are probably not crucial to the alternative perspective or its further progress. The latter will depend on more serious conceptual issues, some of which are the following: 

(i) How come the microstructure of spacetime  exhibits itself indirectly through the 
horizon temperature even at scales much larger than Planck length? I believe this is 
because the event horizon works as some kind of magnifying glass allowing us to probe
trans-Planckian physics \cite{magglass} but this notion needs to be made more precise.

(ii) How does one obtain the expression for entropy in \eq{ent-func-2} from some microscopic model? In particular, such an analysis  --- even with a toy model --- should throw more light on why normals to local patches of null surfaces play such a crucial role as effective degrees of freedom in the long wavelength limit. Of course, such a model
should also determine the expression for $P^{abcd}$ and get the metric tensor and spacetime as derived concepts - a fairly tall order!. (This is somewhat like obtaining theory of elasticity starting from a microscopic model for a solid, which, incidentally, is not a simple task either.)

\ha But what about the ``deep questions'' like, for example,
 the physics near the singularities? Since you get the same field equations as anybody else does, you will have the same solutions, same singularities etc.
 
 \me I told you that I am not doing statistical mechanics (which would be the full quantum theory) of spacetime but only thermodynamics. To answer issues related to singularities etc., one actually needs to discover the statistical mechanics underlying the thermodynamic description I have presented here. We can have another chat, after I figure out the statistical mechanics of the spacetime microstructure!

 \section*{Acknowledgements}
 
 The questions of Harold mostly represent issues raised by several colleagues --- far too numerous to name individually --- in my lectures, discussions etc. I thank all of them for helping me to sharpen the ideas. I also thank A.D. Patel, K. Subramanian, Sudipta Sarkar, Aseem Paranjape, D. Kothawala and  Sunu Engineer for several rounds of discussions over the past many years.

\end{document}

%% file: ray1.pstex_t
\begin{picture}(0,0)%
\includegraphics{ray1.pstex}%
\end{picture}%
\setlength{\unitlength}{3947sp}%
\begingroup\makeatletter\ifx\SetFigFont\undefined%
\gdef\SetFigFont#1#2#3#4#5{%
  \reset@font\fontsize{#1}{#2pt}%
  \fontfamily{#3}\fontseries{#4}\fontshape{#5}%
  \selectfont}%
\fi\endgroup%
\begin{picture}(10802,8702)(600,-8462)
\put(3226,-3886){\makebox(0,0)[lb]{\smash{{\SetFigFont{25}{30.0}{\familydefault}{\mddefault}{\updefault}{\color[rgb]{0,0,0}Arbitrary}%
}}}}
\put(3226,-4216){\makebox(0,0)[lb]{\smash{{\SetFigFont{25}{30.0}{\familydefault}{\mddefault}{\updefault}{\color[rgb]{0,0,0}Event}%
}}}}
\put(5776,-5101){\makebox(0,0)[lb]{\smash{{\SetFigFont{25}{30.0}{\familydefault}{\mddefault}{\updefault}{\color[rgb]{0,0,0}$\mathcal{P}$}%
}}}}
\put(8701,-4936){\makebox(0,0)[lb]{\smash{{\SetFigFont{25}{30.0}{\familydefault}{\mddefault}{\updefault}{\color[rgb]{0,0,0}Local Rinder}%
}}}}
\put(8701,-5266){\makebox(0,0)[lb]{\smash{{\SetFigFont{25}{30.0}{\familydefault}{\mddefault}{\updefault}{\color[rgb]{0,0,0}observer}%
}}}}
\put(5326,-6811){\makebox(0,0)[lb]{\smash{{\SetFigFont{25}{30.0}{\familydefault}{\mddefault}{\updefault}{\color[rgb]{0,0,0}Null rays}%
}}}}
\put(5326,-7141){\makebox(0,0)[lb]{\smash{{\SetFigFont{25}{30.0}{\familydefault}{\mddefault}{\updefault}{\color[rgb]{0,0,0}through $\mathcal{P}$}%
}}}}
\put(1051,-136){\makebox(0,0)[lb]{\smash{{\SetFigFont{25}{30.0}{\familydefault}{\mddefault}{\updefault}{\color[rgb]{0,0,0}$\bar t$}%
}}}}
\put(3151,-1786){\makebox(0,0)[lb]{\smash{{\SetFigFont{25}{30.0}{\familydefault}{\mddefault}{\updefault}{\color[rgb]{0,0,0}$\bar x$}%
}}}}
\end{picture}%

%% file: ray2.pstex_t
\begin{picture}(0,0)%
\includegraphics{ray2.pstex}%
\end{picture}%
\setlength{\unitlength}{3947sp}%
\begingroup\makeatletter\ifx\SetFigFont\undefined%
\gdef\SetFigFont#1#2#3#4#5{%
  \reset@font\fontsize{#1}{#2pt}%
  \fontfamily{#3}\fontseries{#4}\fontshape{#5}%
  \selectfont}%
\fi\endgroup%
\begin{picture}(10802,8702)(600,-8462)
\put(3151,-1786){\makebox(0,0)[lb]{\smash{{\SetFigFont{25}{30.0}{\familydefault}{\mddefault}{\updefault}{\color[rgb]{0,0,0}$X$}%
}}}}
\put(1051,-6211){\makebox(0,0)[lb]{\smash{{\SetFigFont{25}{30.0}{\familydefault}{\mddefault}{\updefault}{\color[rgb]{0,0,0}$\mathcal{R} xx \sim 1$}%
}}}}
\put(9001,-5461){\makebox(0,0)[lb]{\smash{{\SetFigFont{25}{30.0}{\familydefault}{\mddefault}{\updefault}{\color[rgb]{0,0,0}Local Rinder}%
}}}}
\put(8866,-6301){\makebox(0,0)[lb]{\smash{{\SetFigFont{25}{30.0}{\familydefault}{\mddefault}{\updefault}{\color[rgb]{0,0,0}$\kappa^{-1}\ll \mathcal{R}^{-1/2}$}%
}}}}
\put(8986,-5866){\makebox(0,0)[lb]{\smash{{\SetFigFont{25}{30.0}{\familydefault}{\mddefault}{\updefault}{\color[rgb]{0,0,0}observer}%
}}}}
\put(1036,-7051){\makebox(0,0)[lb]{\smash{{\SetFigFont{25}{30.0}{\familydefault}{\mddefault}{\updefault}{\color[rgb]{0,0,0}local inertial frame}%
}}}}
\put(1066,-6631){\makebox(0,0)[lb]{\smash{{\SetFigFont{25}{30.0}{\familydefault}{\mddefault}{\updefault}{\color[rgb]{0,0,0}limit of validity of}%
}}}}
\put(5791,-5266){\makebox(0,0)[lb]{\smash{{\SetFigFont{25}{30.0}{\familydefault}{\mddefault}{\updefault}{\color[rgb]{0,0,0}$\mathcal{P}$}%
}}}}
\put(1126,-136){\makebox(0,0)[lb]{\smash{{\SetFigFont{25}{30.0}{\familydefault}{\mddefault}{\updefault}{\color[rgb]{0,0,0}$T$}%
}}}}
\end{picture}%

%% file: ray3.pstex_t
\begin{picture}(0,0)%
\includegraphics{ray3.pstex}%
\end{picture}%
\setlength{\unitlength}{3947sp}%
\begingroup\makeatletter\ifx\SetFigFont\undefined%
\gdef\SetFigFont#1#2#3#4#5{%
  \reset@font\fontsize{#1}{#2pt}%
  \fontfamily{#3}\fontseries{#4}\fontshape{#5}%
  \selectfont}%
\fi\endgroup%
\begin{picture}(10802,9008)(600,-8768)
\put(6151,-811){\makebox(0,0)[lb]{\smash{{\SetFigFont{25}{30.0}{\familydefault}{\mddefault}{\updefault}{\color[rgb]{0,0,0}$T_E$}%
}}}}
\put(7801,-3661){\makebox(0,0)[lb]{\smash{{\SetFigFont{25}{30.0}{\familydefault}{\mddefault}{\updefault}{\color[rgb]{0,0,0}$x$}%
}}}}
\put(10351,-5236){\makebox(0,0)[lb]{\smash{{\SetFigFont{25}{30.0}{\familydefault}{\mddefault}{\updefault}{\color[rgb]{0,0,0}$X$}%
}}}}
\put(8566,-7786){\makebox(0,0)[lb]{\smash{{\SetFigFont{25}{30.0}{\familydefault}{\mddefault}{\updefault}{\color[rgb]{0,0,0}Local Rinder}%
}}}}
\put(8596,-8206){\makebox(0,0)[lb]{\smash{{\SetFigFont{25}{30.0}{\familydefault}{\mddefault}{\updefault}{\color[rgb]{0,0,0}observer}%
}}}}
\put(8581,-8656){\makebox(0,0)[lb]{\smash{{\SetFigFont{25}{30.0}{\familydefault}{\mddefault}{\updefault}{\color[rgb]{0,0,0}$\kappa^{-1}\ll \mathcal{R}^{-1/2}$}%
}}}}
\put(1516,-1006){\makebox(0,0)[lb]{\smash{{\SetFigFont{25}{30.0}{\familydefault}{\mddefault}{\updefault}{\color[rgb]{0,0,0}$\mathcal{R} xx \sim 1$}%
}}}}
\put(1501,-1861){\makebox(0,0)[lb]{\smash{{\SetFigFont{25}{30.0}{\familydefault}{\mddefault}{\updefault}{\color[rgb]{0,0,0}local inertial frame}%
}}}}
\put(1501,-1411){\makebox(0,0)[lb]{\smash{{\SetFigFont{25}{30.0}{\familydefault}{\mddefault}{\updefault}{\color[rgb]{0,0,0}limit of validity of}%
}}}}
\put(6901,-5026){\makebox(0,0)[lb]{\smash{{\SetFigFont{20}{24.0}{\familydefault}{\mddefault}{\updefault}{\color[rgb]{0,0,0}$\kappa t_E$}%
}}}}
\put(5881,-5071){\makebox(0,0)[lb]{\smash{{\SetFigFont{25}{30.0}{\familydefault}{\mddefault}{\updefault}{\color[rgb]{0,0,0}$\mathcal{P}$}%
}}}}
\end{picture}%